\documentclass{PoS}

\usepackage{amsmath}
\usepackage{amssymb}

\usepackage{cite}

\usepackage{epsfig}

\newcommand{\be}{\begin{equation}}
\newcommand{\ee}{\end{equation}}
\newcommand{\bea}{\begin{eqnarray}}
\newcommand{\eea}{\end{eqnarray}}
\newcommand{\nn}{\nonumber}
\newcommand{\bra}{\langle}
\newcommand{\ket}{\rangle}
\newcommand{\Tr}{\mbox{Tr}\,}
\newcommand{\hm}{\hspace*{-0.2cm}}
\newcommand{\rmI}{{\rm I}}
\newcommand{\half}{\frac{1}{2}}
\newcommand{\re}{{\rm Re}}
\newcommand{\im}{{\rm Im}}

\title{Complex Langevin dynamics and other approaches at finite chemical potential }

\ShortTitle{Complex Langevin dynamics and other approaches at finite chemical
potential }

\author{\speaker{Gert Aarts}%
        \addtocounter{footnote}{2} \thanks{Plenary talk.}\\
        Department of Physics, College of Science, 
        Swansea University, Swansea, United Kingdom\\
        E-mail: \email{g.aarts@swan.ac.uk}}

\abstract{I review the presence of the sign problem in lattice QCD at nonzero baryon density and its relation with the overlap and Silver Blaze problems. I then discuss progress in some cases where the sign problem can be handled, either because the sign problem is absent or because it is milder than in full QCD. Some time is spent on effective three-dimensional models, which can be treated with a variety of methods. I conclude with a discussion of the applicability of complex Langevin dynamics at nonzero density.
}

\FullConference{The 30th International Symposium on Lattice Field Theory\\
                 June 24 -- 29,  2012\\
                 Cairns, Australia}

\begin{document}

\section{Introduction}

\begin{figure}[h]
\vspace*{0.1cm}
 \begin{center}
  \includegraphics[height=3.9cm]{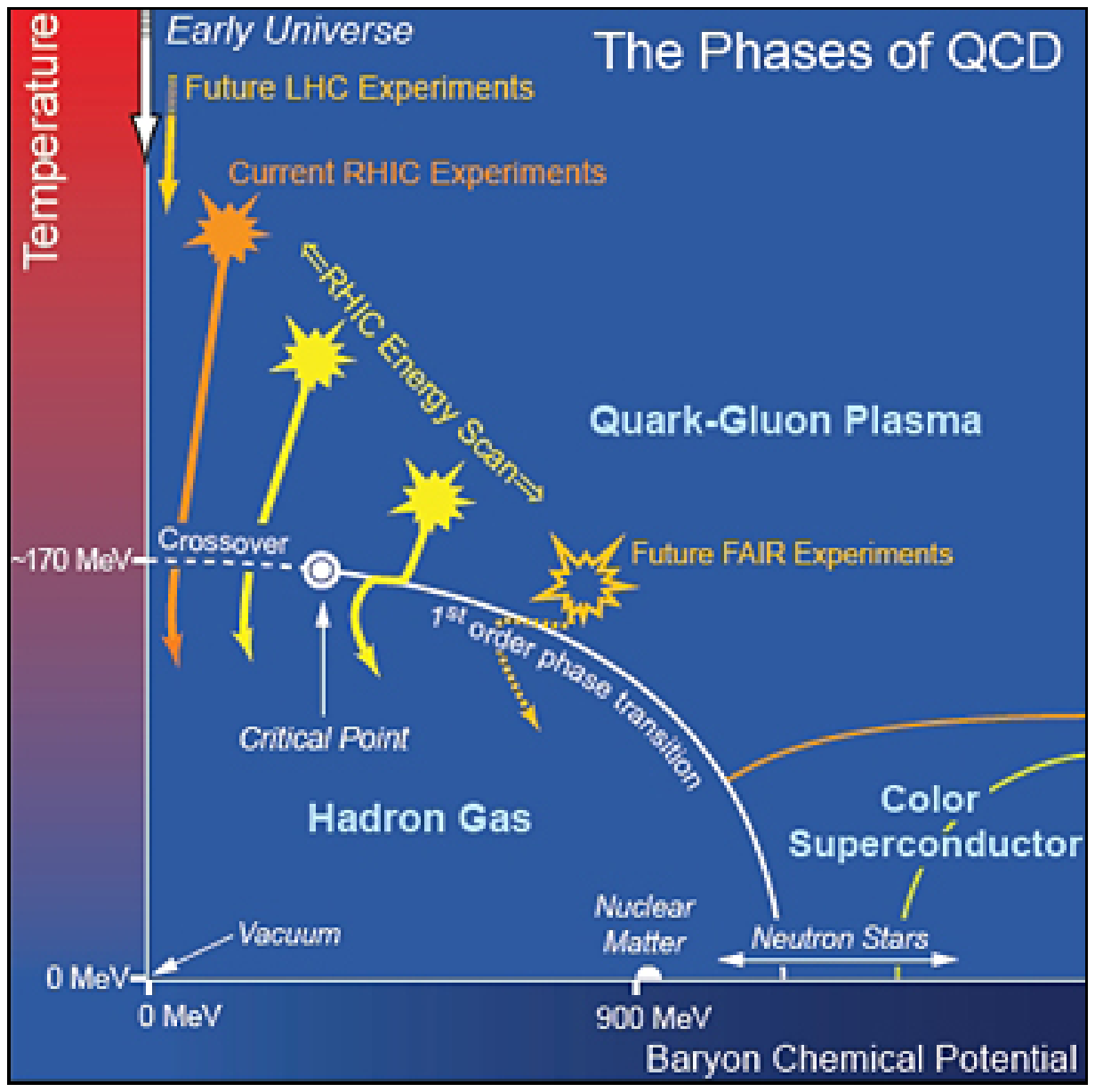}
   \includegraphics[height=4cm]{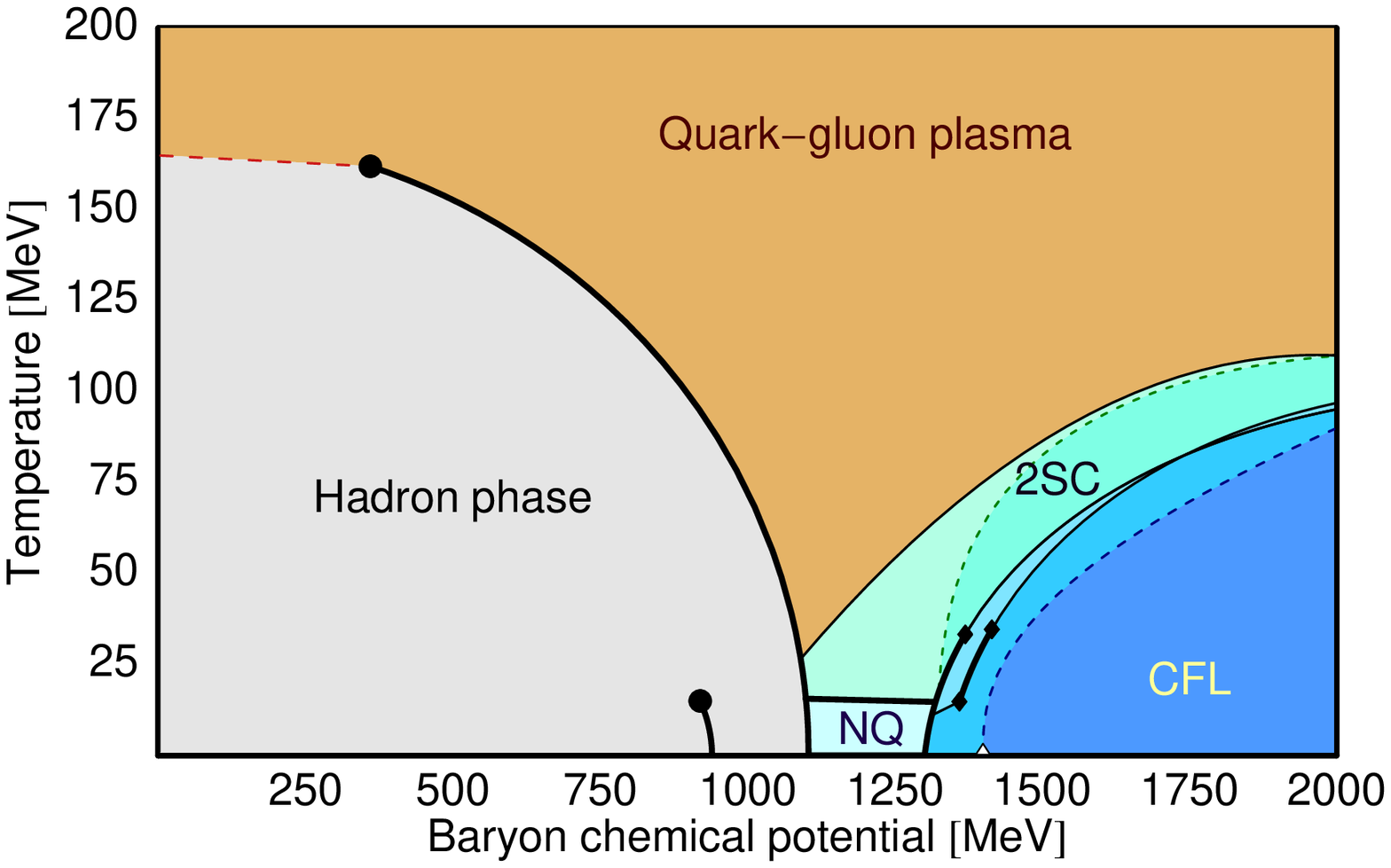}
   \includegraphics[height=4cm]{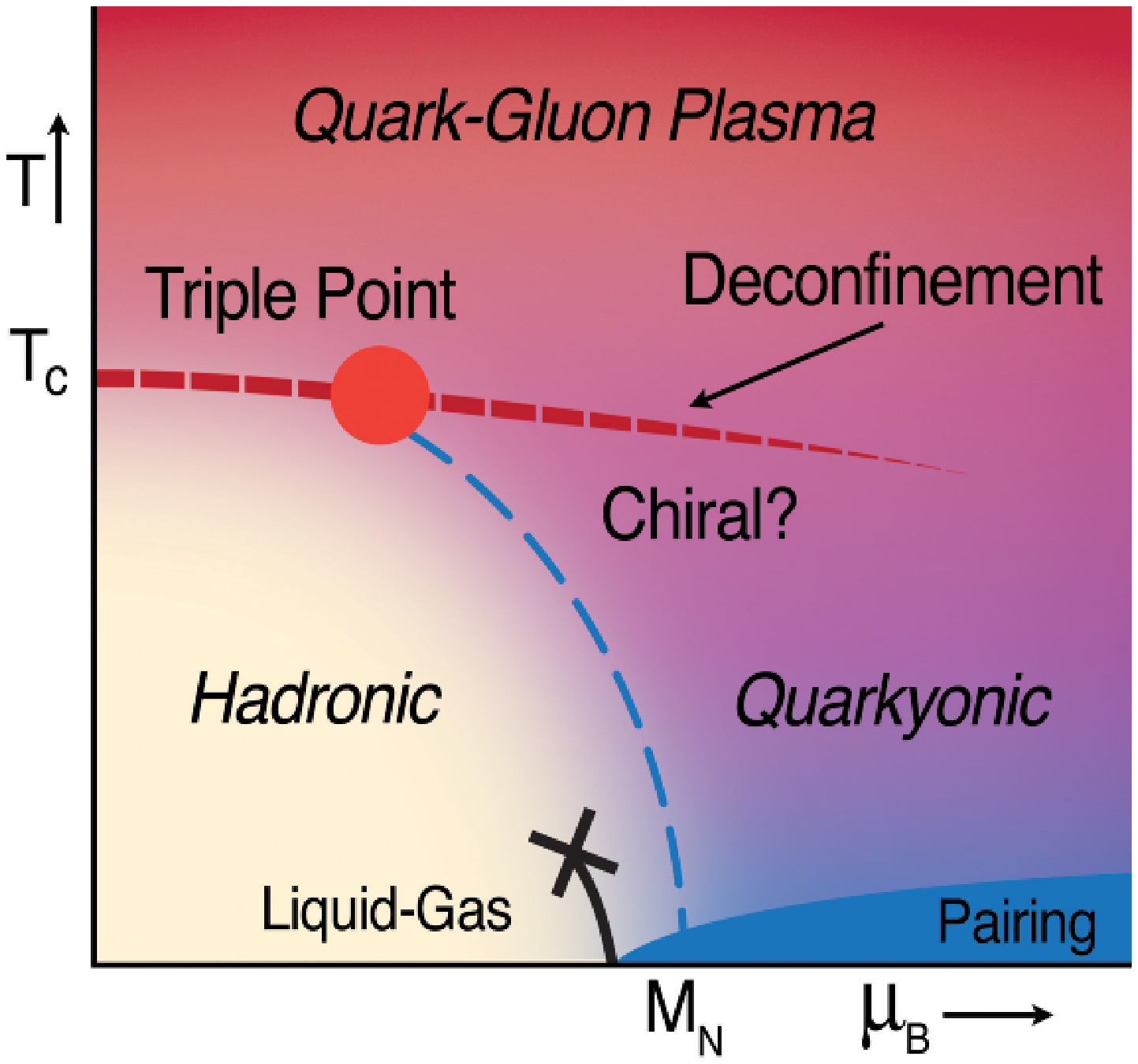}
 \caption{Some impressions of the QCD phase diagram.}
\label{fig:pd}
\end{center}
\end{figure}

The phase structure of strongly interacting matter, i.e.\ the QCD phase diagram, has not yet been determined from first principles. The reason is of course well-known: at nonzero baryon chemical potential the fermion determinant is complex and this prevents the use of the standard numerical methods commonly used in lattice QCD simulations.
As a result there is no agreed QCD phase diagram in the plane spanned by the temperature $T$ and baryon chemical potential $\mu_B$ (or quark chemical potential $\mu_q=\mu_B/3$). Some colourful impressions   appearing close to the top of a Google search (yielding around 19000 hits) are shown in fig.\ \ref{fig:pd}.

What is understood is the transition along the temperature axis for vanishing and small chemical potential. This is highly relevant since the crossover from the confined hadronic phase to the deconfined quark-gluon plasma phase is currently being probed by heavy-ion collisions at RHIC and the LHC. The theoretical status of the phase diagram in this region was reviewed by Maria-Paola Lombardo at this conference \cite{Lombardo:2012ix}.
At larger chemical potential the sign problem prohibits the use of well-established methods.  In this contribution I will first discuss  the close connection between the sign, the overlap and the Silver Blaze problems and how they appear in numerical simulations. Subsequently I will present results in cases where simulations {\em are} possible, either because the sign problem is absent or because it is much milder than in full QCD.  In the third and final part of this review I will motivate why it makes sense to go into the complex plane and discuss the applicability of complex Langevin dynamics.  I will not review the evidence in favour or against the QCD critical endpoint, and quarkyonic and colour-superconducting phases.  Also, ``standard methods'' at small $\mu/T$, such as reweighting, Taylor series, analytical continuation, histograms or the canonical ensemble, have been partly reviewed at previous gatherings \cite{deForcrand:2010ys,arXiv:1101.0109,Levkova:2012jd}.

\section{Sign, overlap and Silver Blaze problems}

In the standard formulation of lattice QCD, the quarks are integrated out analytically, yielding the fermion determinant, while the functional integral over the gluonic variables in the partition function remains to be done, i.e.
\bea
\nn
Z  &=& \int DU D\bar\psi D\psi \, e^{ -S_{\rm YM}(U)-S_{\rm F}(U;\mu)} \\
&=&  \int DU\, e^{-S_{\rm YM}(U)} \det\, M(U;\mu).
\eea
Here $M$ is the fermion matrix, appearing in the fermionic action as
\be
S_{\rm F}(U;\mu) = -\int d^4x\, \bar\psi\, M(U;\mu)\,\psi.
\ee
The chemical potential is introduced in the usual way, as the fourth component of an imaginary vector potential 
\cite{Hasenfratz:1983ba}; the dependence of $M$ on the gauge links $U$ will no longer be indicated.
 If the combined Boltzmann weight, from the Yang-Mills action and from the fermion determinant, is real and nonnegative, importance sampling can be used to evaluate the remaining bosonic  high-dimensional integral numerically. This is the usual situation in lattice QCD at zero and nonzero temperature.

At nonzero baryon density, or at nonzero baryon chemical potential in the grand-canonical ensemble, this is however not the case. Due to the lack of $\gamma_5$-hermiticity,
\be
\label{eq:gamma5}
\gamma_5M(\mu)\gamma_5 = M^\dagger(-\mu^*) \neq  M^\dagger(\mu) ,
\ee
the determinant is complex and satisfies 
\be
[\det\, M(\mu)]^* = \det\, M(-\mu^*).
\ee
The problems associated with a complex or nonpositive weight are collectively referred to as the {\em sign problem}, even though a complex phase problem would be more accurate.
It is important to realize that the complexity does not arise from the Grassmann nature of the quark fields. After all, at vanishing chemical potential, the determinant is real. It is therefore not a {\em fermion sign problem} per se.  In fact, the sign problem also appears in bosonic theories at nonzero chemical potential, in which the complex action enjoys the same symmetry properties as the determinant, i.e.\ $S^*(\mu)=S(-\mu^*)$. In all these cases the complexity is absent for purely imaginary chemical potentials.

The simplest approach would be to write 
\be
\det\,  M  = | \det\,  M |e^{i\theta},
\ee
and ignore the phase factor at first instance. Simulations in the resulting phase-quenched (pq) theory are straightforward in principle, since they do not suffer from the sign problem. The phase factor can be incorporated later via reweighting, 
\be
\bra O\ket = \frac{\int DU\, e^{-S_{\rm YM}} \det\, M \, O }{\int DU\, e^{-S_{\rm YM}} \det\, M}
= \frac{\int DU\, e^{-S_{\rm YM}} |\det\, M| e^{i\theta} \, O }{\int DU\, e^{-S_{\rm YM}} |\det\, M| e^{i\theta}}
= \frac{\bra e^{i\theta} \, O \ket_{\rm pq}}{\bra e^{i\theta}\ket_{\rm pq}},
\ee
so that observables are recovered from ratios in the phase-quenched theory.

However, when the full and the phase-quenched theory differ, this is bound to lead to an {\em overlap problem}: the generated phase-quenched ensemble has little in common with the desired ensemble in the original theory. This is easily understood mathematically: the average phase factor $\bra e^{i\theta}\ket_{\rm pq}$ in the phase-quenched theory is nothing else than the ratio of two partition functions, 
\be 
\bra e^{i\theta}\ket_{\rm pq} =  \frac{\int DU\, e^{-S_{\rm YM}} |\det\, M| e^{i\theta} }{\int DU\, e^{-S_{\rm YM}} |\det\, M|} = \frac{Z}{Z_{\rm pq}} \equiv  e^{-V\Delta f/T},
\quad\quad\quad
\Delta f = f-f_{\rm pq},
\ee
where $f=-(T/V)\ln Z$ is the free energy density in the full  theory, $V$ the three-dimensional volume and $T$ the temperature (and similar for $f_{\rm pq}$ in the phase-quenched theory).
  When $\Delta f\neq 0$, it follows that the average phase factor will go to zero in an exponential fashion when the thermodynamic limit is taken and the temperature is reduced. In lattice simulations at fixed lattice spacing, this scaling is with the four-volume, $V/T\sim \Omega \equiv N_s^3N_\tau$. This behaviour of the phase factor, and the related worsening of the signal-to-noise ratio, is unavoidable and limits the system sizes that can be addressed with this approach. If this happens, the sign problem is {\em severe}.  
   Importantly, $\Delta f$ depends on the theory  or model under consideration  and in some cases, such as in QCD in the strong-coupling limit \cite{deForcrand:2009dh} or in certain effective three-dimensional models (see below), $\Delta f$ is smaller than perhaps expected and the sign problem can be rather {\em mild}. Simulations are then possible for reasonably large -- but still finite -- system sizes.

  There is also a physical reason for the presence of the sign problem. This is especially clear at zero temperature, in the case that the transition from vacuum to a high-density state (the {\em onset}) occurs at a lower critical chemical potential in the phase-quenched theory than in the original theory, i.e. $\mu_c^{\rm pq}<\mu_c$. The complex phase should then produce excessive cancellations to wash out the early onset in the phase-quenched theory.  
  This elimination has to be perfect, since in the original theory at zero temperature thermodynamic quantities are independent of $\mu$ as long as $\mu<\mu_c$. 
  This has been named the {\em Silver Blaze problem} \cite{Cohen:2003kd,silverblaze}: while every numerically generated configuration in the ensemble depends on the chemical potential, this chemical potential dependence should cancel exactly in thermodynamic observables, when evaluated in the entire ensemble. The mechanism behind this might be highly nontrivial.
The region where $\mu_c^{\rm pq}<\mu<\mu_c$ is referred to as the {\em Silver Blaze region}. 
 
A convenient example to illustrate this is QCD with two flavours of equal mass, and hence fermion determinant $[\det M(\mu)]^2$ ($\mu$ is the quark chemical potential). The determinant in the phase-quenched theory is different and reads
 $|\det M(\mu)|^2 =  \det M(\mu)\det M(-\mu)$, i.e.\ the phase-quenched theory corresponds to a theory with nonzero isospin chemical potential \cite{Son:2000xc}. The onsets occur at different critical chemical potentials: in the isospin theory one finds condensation of pions at $\mu_c^{\rm pq}=m_\pi/2$, while at finite quark number one expects onset to nuclear matter at $\mu_c\sim m_N/3$ ($-$ binding energy), where $m_N$ denotes the nucleon mass. The Silver Blaze region is therefore the region with $m_\pi/2<\mu\lesssim m_N/3$.
It has indeed been shown that in the Silver Blaze region the $\mu$-dependent spectral density of the Dirac operator is complex and highly oscillatory, with a diverging amplitude $\exp(\Omega)$ and period $1/\Omega$, to ensure $\mu$-independence of the chiral condensate and baryon number
(OSV mechanism) \cite{Osborn:2005ss,Osborn:2008jp}. Only when the integration is done carefully can the desired cancellations take place. This important insight indicates some of the difficulties one may encounter when approximate methods are used.

The discussion above refers to the case that the sign problem has not been resolved. However, in some cases the sign problem can be eliminated by a reformulation in other degrees of freedom or by the use of complex Langevin dynamics. In those cases, the distinction between mild and severe sign problems is no longer necessary. Similarly, there is no longer a Silver Blaze problem. However, it is still useful to study the approach in particular in the Silver Blaze region, since this provides a stringent test on whether the sign problem has indeed been eliminated correctly.

After setting the stage, I will continue with some selected results in theories without a sign problem and subsequently discuss theories with a milder sign problem, in particular effective three-dimensional models. Finally, I will argue that  one has to take the complexity at nonzero chemical potential seriously and embark on a journey into the complex plane.

\section{Theories without a sign problem}
\label{sec:wo}

The prime example of a nonabelian gauge theory without a sign problem is two-colour QCD or QC$_2$D: QCD with gauge group SU(2) and quarks transforming in the fundamental representation. In this theory, there is an additional Pauli--G\"ursey symmetry \cite{Hands:1999md},
\be
K M(\mu) K^{-1} = M^*(\mu), \quad\quad\quad K\equiv C\gamma_5\tau_2,
\ee
with $C$ the charge-conjugation matrix. While $\gamma_5$-hermiticity  is still absent at nonzero chemical potential, see Eq.\ (\ref{eq:gamma5}), the Pauli--G\"ursey symmetry nevertheless leads to a real determinant, even when $\mu\neq 0$.  For two flavours and in the presence of diquark sources, the determinant can then be written in a manifestly positive manner and lattice simulations can be carried out using conventional algorithms.

\begin{figure}[h]
 \begin{center}
  \includegraphics[height=7cm]{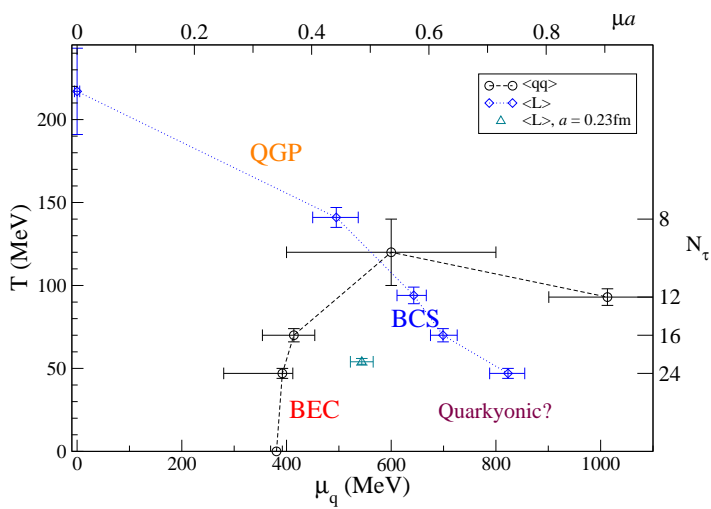}
 \caption{Tentative phase diagram in QC$_2$D \cite{Cotter:2012mb}.}
 \end{center}
\label{fig:qc2d}
\end{figure}

Lattice QC$_2$D at nonzero chemical potential has been studied already for some time. Recent progress concerns the quarkonium spectrum at low temperature  and large chemical potential \cite{Hands:2012yy}, a detailed analysis of singular values of the Dirac operator, determined by $D^\dagger(\mu)D(\mu)\psi_n=\xi_n^2\psi(n)$ \cite{Kanazawa:2011tt} (in contrast to the eigenvalues of the Dirac operator, the singular values remain real and nonnegative),
and the phase diagram at nonzero temperature and density \cite{Cotter:2012mb}. The latter is shown in Fig.~\ref{fig:qc2d}, with the axes labelled both in MeV and in lattice units. Besides the usual confined and quark-gluon plasma phases, one encounters a low-temperature phase with a nonzero diquark condensate $\bra qq\ket$ as well as a quarkyonic phase. While the transition from vacuum to a diquark-condensed phase should not appear in QCD with three colours, in QC$_2$D it is indeed the expected behaviour, since baryons are bosons and the lightest baryon is degenerate with the pion.
The evidence for the quarkyonic phase is discussed in Ref.\ \cite{Hands:2010gd}.

However, QC$_2$D is not the only theory with a real and positive determinant at nonzero chemical potential. Another interesting example is given by using as gauge group the exceptional group G$_2$ \cite{Maas:2012wr}. Unlike QC$_2$D, this theory has bosonic ($qq$) {\em and} fermionic ($qqq$) baryons, as well as mesonic ($\bar qq$), hybrid ($qggg$) and multi-quark bound states. Thanks to an extended Pauli--G\"ursey symmetry, the determinant is real and positive, even for a single flavour at nonzero $\mu$. 

\begin{figure}[h]
 \begin{center}
  \includegraphics[height=3.8cm]{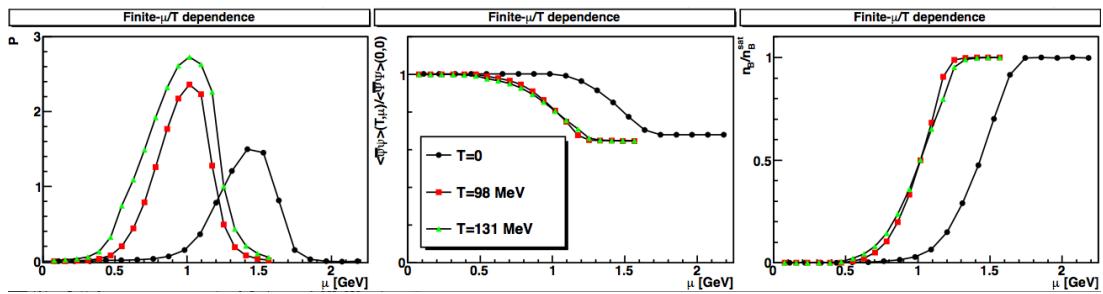}
 \caption{Gauge group G$_2$: chemical potential dependence of the Polyakov loop (left), the normalized chiral condensate (middle) and the normalized baryon density (right) for three temperatures  \cite{Maas:2012wr}.}
\label{fig:G2}
 \end{center}
 \end{figure}

Some first results are presented in Fig.\ \ref{fig:G2}, where the Polyakov loop, the chiral condensate and the baryon density are shown as a function of chemical potential for three temperatures. The onset is presumably due to a diquark condensate, as in QC$_2$D, but there are indications for a second rise in density at larger chemical potential (not shown). To settle this will require precise calculations of the spectrum at zero temperature. The drop of the Polyakov loop at larger $\mu$ is most likely a saturation effect, see e.g.\ the behaviour of the density on the right, and is hence a lattice artefact. Nevertheless, these first results are quite encouraging and clearly deserve further effort.

\section{Theories with a milder sign problem}
\label{sec:mild}

From the perspective of lattice QCD, effective lower-dimensional field theories or spin models provide a useful entry into the study of the QCD phase diagram for a number of reasons. First of all, it has become clear that these models may have a milder sign problem and are therefore more amenable to standard lattice techniques, in combination with reweighting. Secondly, sometimes they can be reformulated using different (or dual) degrees of freedom, such as in world line or flux representations, in which the sign problem is manifestly
absent. And finally, some of these models can be solved numerically with complex Langevin dynamics, providing a completely independent approach. One role these models therefore have is to provide a useful testbed for new methods.

Ideally, the effective models are constructed in such a way that a detailed mapping between the  couplings appearing in those and in full QCD is available: the insight obtained can then be translated back to QCD. To illustrate this, I start with the simplest three-dimensional  SU(3) spin model as a guide. The action, appearing as $\exp(-S)$, consists of two parts  and is written as
\be
S=S_B+S_F, 
\ee
with
\be
\label{eq:SU3}
 S_B = -\beta\sum_{\bra x y\ket} \left[ P_x P_y^* + P_x^* P_y\right],
 \quad\quad\quad
 S_F = -h\sum_x\left[ e^\mu P_x+e^{-\mu} P_x^*\right].
 \ee
 Here the degrees of freedom are effective Polyakov loops, $P_x=\Tr U_x$, $P_x^*=\Tr U_x^\dagger$, where the $U_x$'s are SU(3) matrices, living on a three-dimensional lattice. The QCD gauge action has been replaced by a nearest-neighbour  interaction, while static (anti)quarks are represented by (conjugate) Polyakov loops, weighted with the chemical potential to introduce an imbalance. The action is indeed complex and $S^*(\mu)=S(-\mu^*)$. The effective parameters can be mapped to the parameters in four-dimensional lattice QCD: at leading order in a combined strong-coupling and hopping expansion, one finds that $\beta\sim (\beta_{\rm 4D}/18)^{N_\tau}$, $h=(2\kappa)^{N_\tau}$, with $\kappa$ the Wilson hopping parameter, and $\mu$ corresponds to $\mu/T$ in four dimensions. I come back to this below.

 This model is of course already quite old: it was studied in a mean-field approximation nearly 30 years ago \cite{Banks:1983me} and it was one of the earliest models studied with complex Langevin dynamics \cite{KW,BGS}.
Nevertheless, it has seen a recent revival for a  number of reasons: it can be reformulated as a flux model without a sign problem \cite{arXiv:1104.2503}; the applicability of complex Langevin has been studied in detail \cite{Aarts:2011zn}; a renewed mean-field analysis is available \cite{Greensite:2012xv}; and in fact it is part of a whole family of high-order strong-coupling models, which have been studied in a systematic fashion in recent years \cite{Fromm:2011qi}.

Let us start with the flux representation, in which the sign problem is absent \cite{arXiv:1104.2503}. In this approach the Boltzmann weight is expanded, as in a classical high-temperature expansion, to all orders and the partition function takes the form of infinite sums of integrals over powers of (conjugate) Polyakov loops at each site $x$, i.e.,
\be
I(n_x, \bar n_x) = \int_{\rm SU(3)} dU_x\, \left(\Tr U_x\right)^{n_x} \left(\Tr U_x^\dagger\right)^{\bar n_x}.
\ee
 The crucial next step is to perform these single-site SU(3) integrals, which is possible in this case. The result is a monomer-dimer system with constraints, since $I(n_x, \bar n_x)$ is only nonzero provided $\left(n_x-\bar n_x\right)$ mod 3 $=0$. The important finding is that all the nonzero weights that appear in this representation are real and positive, even when $\mu\neq 0$. The resulting model can then be solved with importance sampling or a worm algorithm \cite{Mercado:2012ue}.

\begin{figure}[h]
 \begin{center}
    \includegraphics[height=4.8cm]{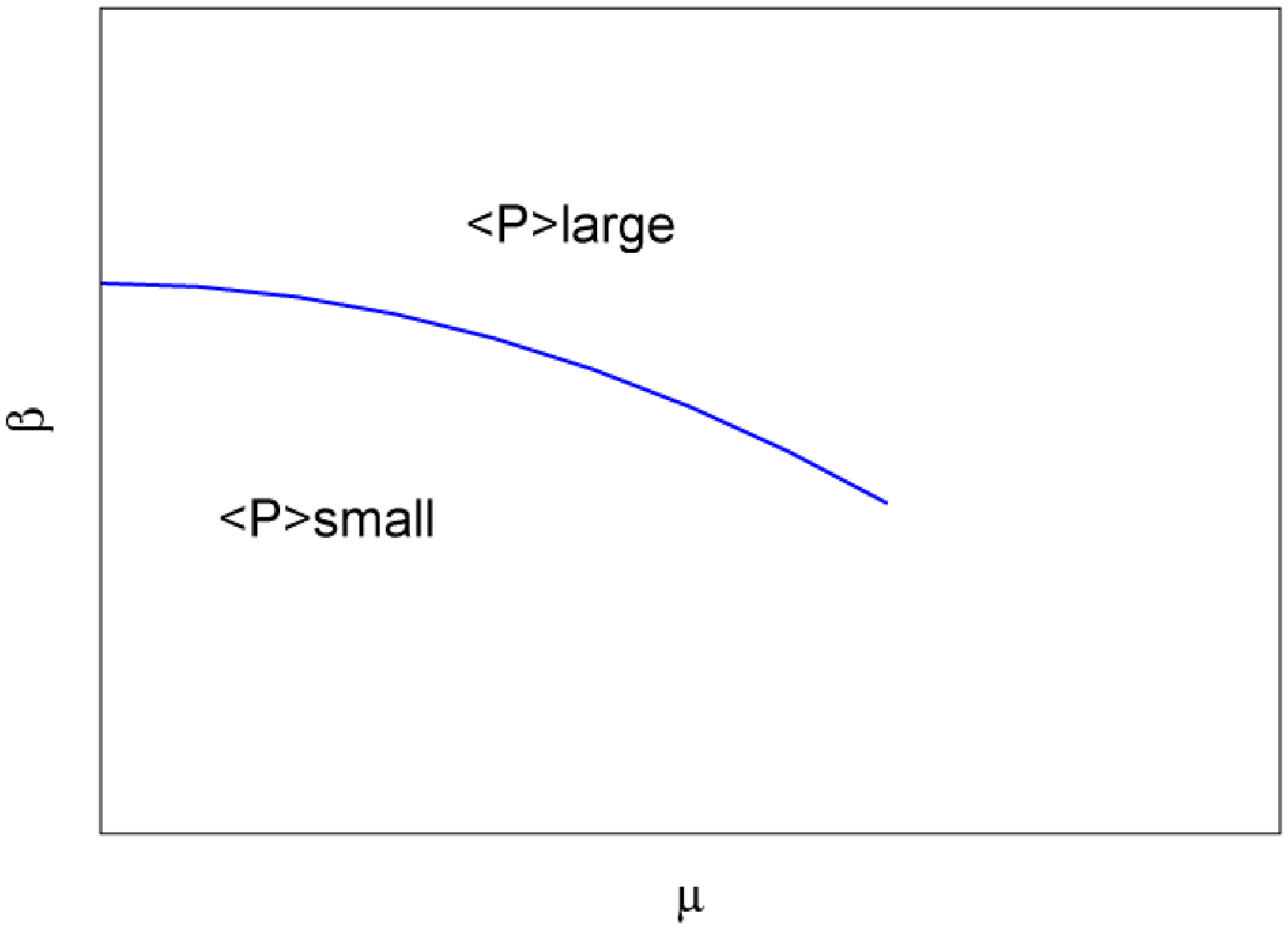} 
    $\quad$
  \includegraphics[height=5cm]{Gattringer-phase.eps}
 \caption{SU(3) spin model: expected phase structure (left) and the result from simulations in the flux representation (right) \cite{Mercado:2012ue}.}
\label{fig:Gat}
 \end{center}
\end{figure}

This theory is expected to have a phase diagram as in Fig.\ \ref{fig:Gat} (left). For small fermion coupling $h$, a quasi-confined or disordered phase (with small $\bra P\ket$, recall that the centre-symmetry is explicitly broken by the fermionic contribution)  and a deconfined  or ordered  phase (with large $\bra P\ket$) are separated by a first-order phase boundary in the $\beta$ -- $\mu$ plane. This line ends in a critical endpoint after which only a crossover remains. The simulations \cite{Mercado:2012ue} confirm this, see Fig.\ \ref{fig:Gat} (right), where phase boundaries are shown for various values of $h$. Increasing $h$, which corresponds to reducing the quark mass, lowers the value of  $\beta_c$ and weakens the transition, as does increasing $\mu$. Note that this is indeed the expected behaviour in full QCD with heavy quarks, as will be discussed below.

The SU(3) spin model (\ref{eq:SU3}) is in fact part of a whole family of effective Polyakov loop models constructed recently \cite{arXiv:1010.0951,Fromm:2011qi,Fromm:2012eb}.
 Systematically integrating out the spatial links in lattice QCD using a strong-coupling expansion while leaving the temporal links untouched  yields effective three-dimensional actions with more and more Polyakov loop interaction terms. For pure SU(3) gauge theory, these take the form \cite{arXiv:1010.0951}
\be
S = -\lambda_1\sum_{\bra xy\ket}  \left[ P_x P_y^* + P_x^* P_y\right] 
 - \lambda_2 \sum_{[ xy ]}  \left[ P_x P_y^* + P_x^* P_y\right] + \mbox{(higher-order representations)} +\ldots,
 \ee
where nearest and next-to-nearest neighbour terms are indicated explicitly. Within each term, certain strong-coupling contributions can be resummed to all orders, using e.g.
\be
S = -\lambda_1\sum_{\bra xy\ket}  \left[ P_x P_y^* + P_x^* P_y\right] + \ldots
= 
-\sum_{\bra xy\ket}  \ln\left[ 1+\lambda_1 \left(P_x P_y^* + P_x^* P_y\right) \right] +\ldots.
\ee
Finally, the dependence of the couplings $\lambda_i$ on $\beta$ and $N_\tau$ is known to quite high order; for instance for $N_\tau=4$ one finds \cite{arXiv:1010.0951}
\bea
\lambda_1(u,N_\tau=4) =&&\hm u^{4}\exp\left[ 4\left(4u^4+12u^5-14u^6+\ldots+
\frac{1035317}{5120}u^{10} +\ldots\right)\right], \\
\lambda_2(u,N_\tau=4) = &&\hm  u^8\left[ 12u^2+26u^4+364u^6+\ldots\right],
\eea
where $u=u(\beta) = a_f(\beta) =  \beta/18+\dots$, as it appears in the character expansion.
By including higher-order terms, the approach can be systematically improved and errors due to the truncation can be  estimated. 

Quarks are taken into account via a hopping expansion; for heavy (static) quarks this yields the well-known contribution 
\cite{Stamatescu:1980br,Bender:1992gn,Blum:1995cb,Aarts:2008rr}
\be
\label{eqSFhop}
S_F\sim -\sum_x\ln\det\left[ \left(1+he^{\mu/T}W_x\right)^2\left(1+he^{-\mu/T}W_x^\dagger\right)^2\right], 
\ee 
where $W_x$ is the untraced Polyakov loop, $h=h(\kappa, N_\tau)$ and the remaining determinant is in colour space. At lowest order in $h$ and $\kappa$, with $h=(2\kappa)^{N_\tau}$, an expansion in $h$ yields the fermionic contribution given in Eq.\ (\ref{eq:SU3}). An important distinction between the two expressions is that the action in Eq.\ (\ref{eqSFhop}) is still a determinant and hence incorporates Fermi-Dirac statistics and saturation on the lattice, while these properties are lost in Eq.\ (\ref{eq:SU3}). This can e.g.\ be  seen by writing down the expression for the density $\bra n\ket = (T/V)\partial \ln Z/\partial \mu$, using the SU(3) identity $\det \left(1+c U\right) = 1+ c\Tr U +c^2 \Tr U^\dagger +c^3$.

\begin{figure}[t]
 \begin{center}
   \includegraphics[height=5cm]{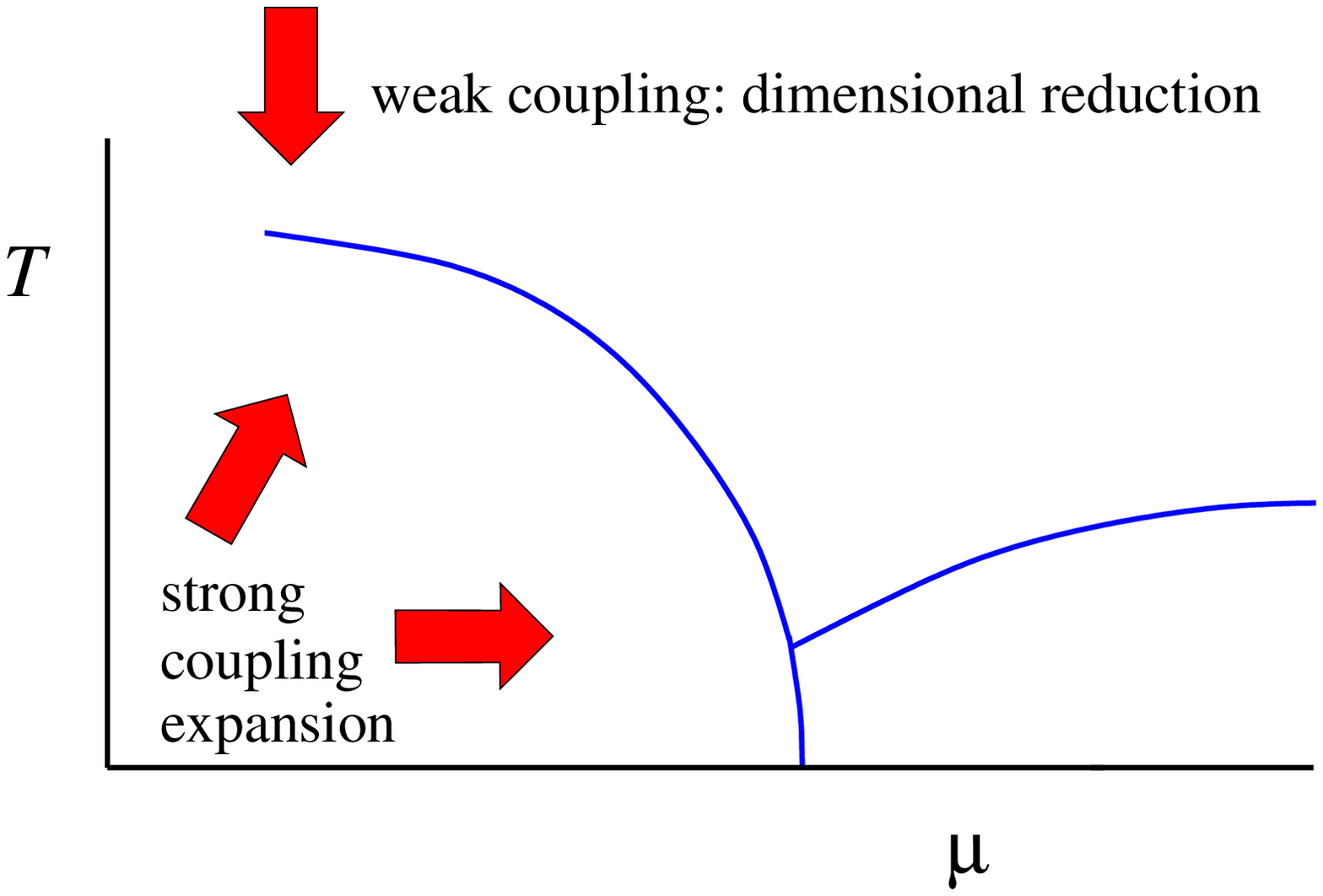}
     $\quad$
  \includegraphics[height=5cm]{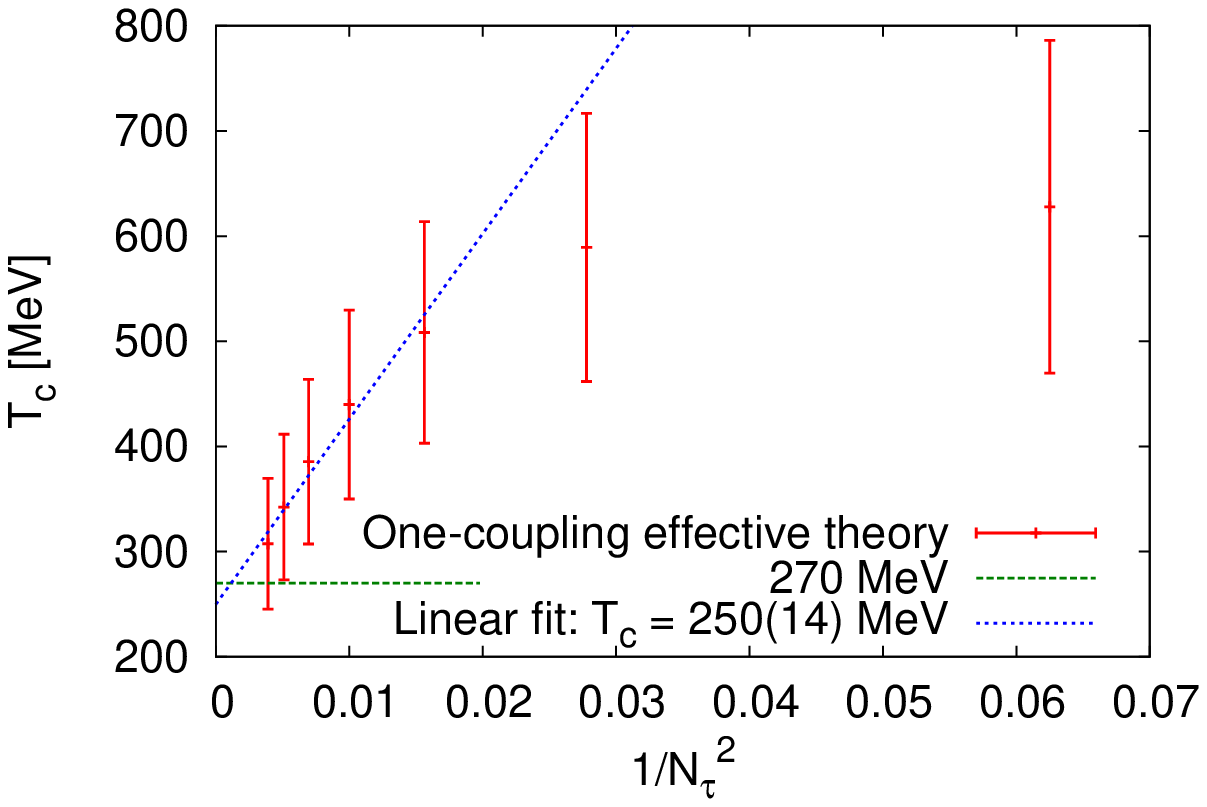}
 \caption{Validity of strong-coupling models, as complementary to dimensional reduction (left). The critical temperature in SU(3) gauge theory from a one-coupling effective theory (right) \cite{Fromm:2011qi}.}
\label{fig:OP}
\end{center}
\end{figure}

It is clear that the approach sketched here goes considerably beyond the simplest SU(3) spin model discussed first and one may ask the question whether quantitative results can be obtained. In order to answer this, it may  for a moment be advantageous to forget about the strong-coupling origin and view the setup instead as a generic approach to construct effective models, whose validity may be justified afterwards. In some sense it can be viewed as complementary to dimensional reduction at high temperature and weak coupling. This is illustrated in Fig.\ \ref{fig:OP} (left). While dimensional reduction, formulated in terms of quarks and gluons, is formally only valid at weak coupling and hence (very) high temperature, it is well-known that it can be applicable down to 1.5--2$T_c$ \cite{Laine:2003ay}. Similarly, while a combined strong-coupling/hopping expansion,  formulated in terms of colour singlets and meson/baryon degrees of freedom, is formally valid in what appears to be an unphysical regime at finite lattice spacing (i.e.\ not in the continuum limit), one may still test its validity. 
This is illustrated in Fig.\ \ref{fig:OP} (right) in pure SU(3) gauge theory. Here the critical temperature from  four-dimensional lattice simulations (270 MeV) is compared with the one obtained from an effective three-dimensional theory with one coupling only, $\lambda_1(\beta, N_\tau)$. Once the critical coupling 
$\lambda_{1,c}(\beta, N_\tau)$ is determined, the results can be mapped back to four dimensions to find $\beta_c(N_\tau)$. After setting the scale, this yields Fig.\ \ref{fig:OP} (right), where all the data points come from a single simulation in the one-parameter effective theory. By comparing truncation effects, an error can be assigned as well.

\begin{figure}[t]
 \begin{center}
  \includegraphics[height=5cm]{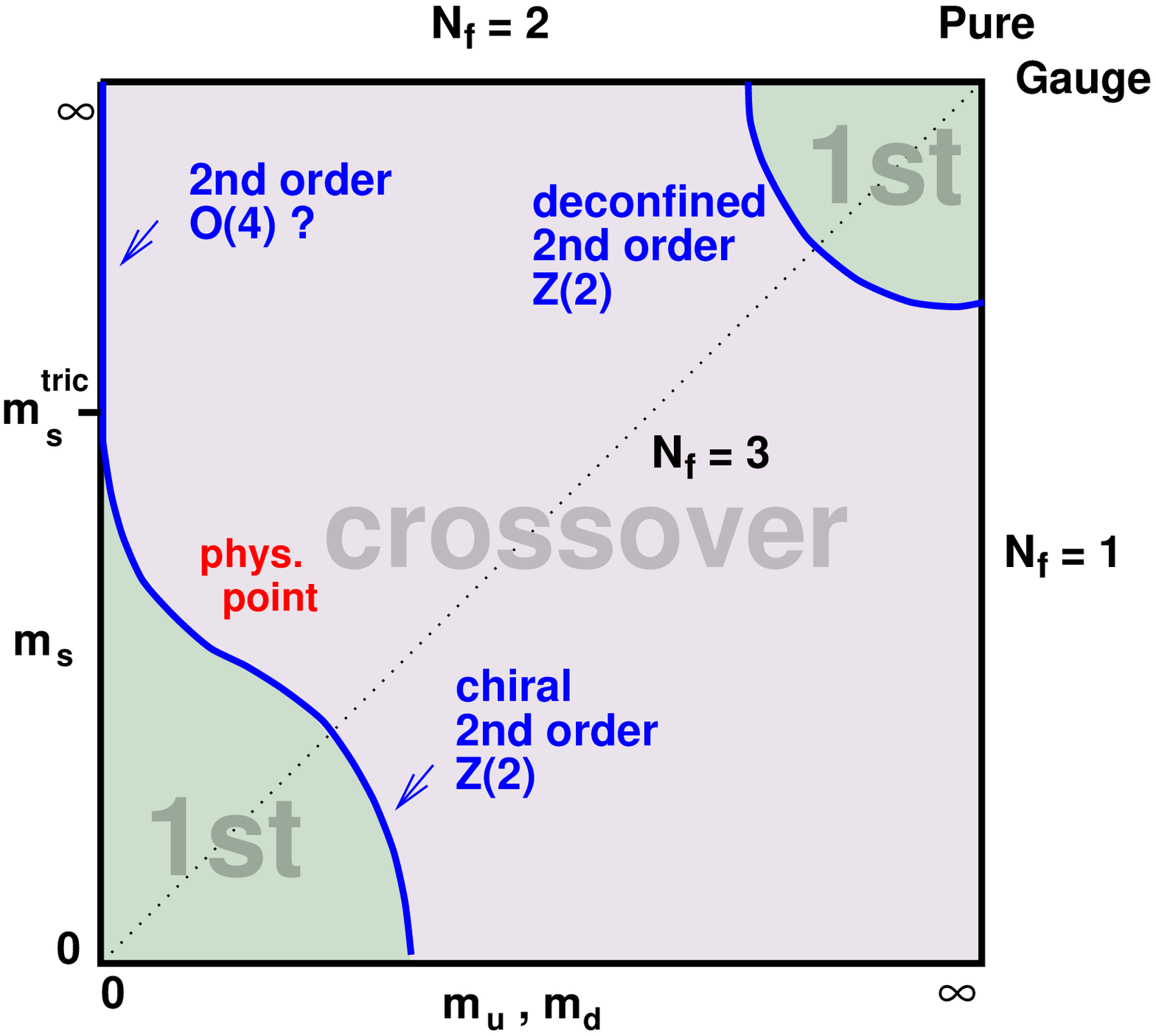} 
  $\quad\quad$
  \includegraphics[height=5cm]{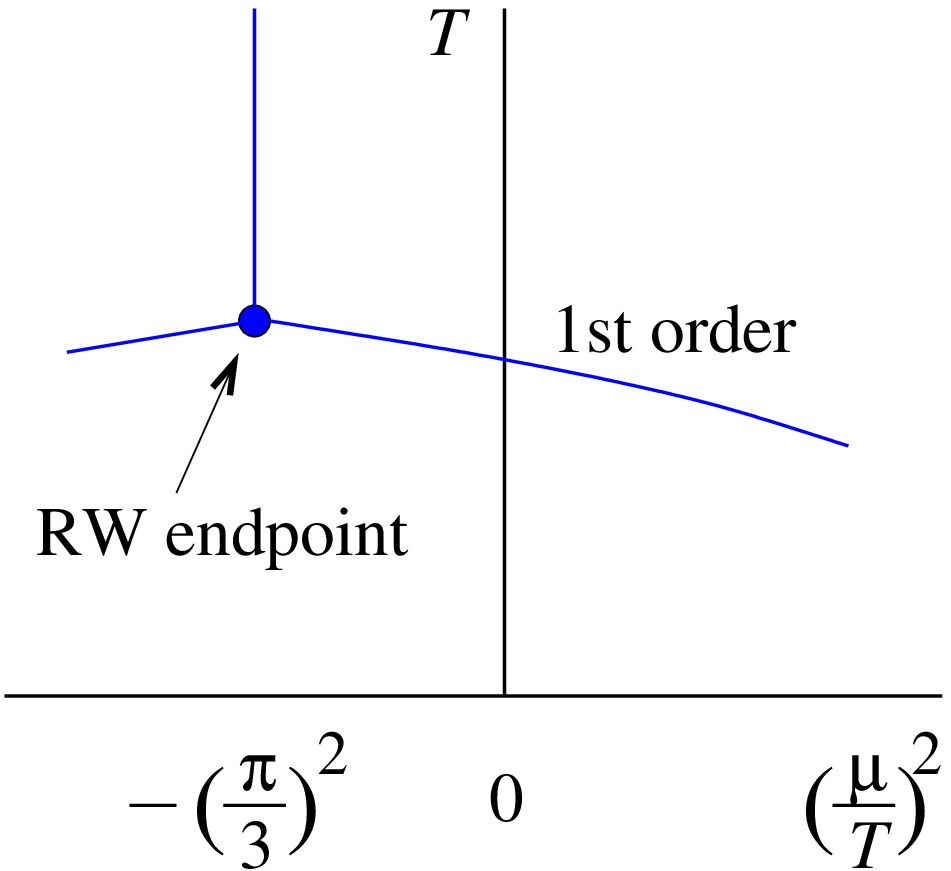}
 \caption{Columbia plot: expected quark mass dependence of the order of the thermal deconfinement transition at vanishing chemical potential (left) and the extension of the QCD phase diagram to nonzero chemical potential in the case of heavy quarks (right).}
\label{fig:Col}
\end{center}
\end{figure}

\begin{figure}[t]
 \begin{center}
  \includegraphics[height=5cm]{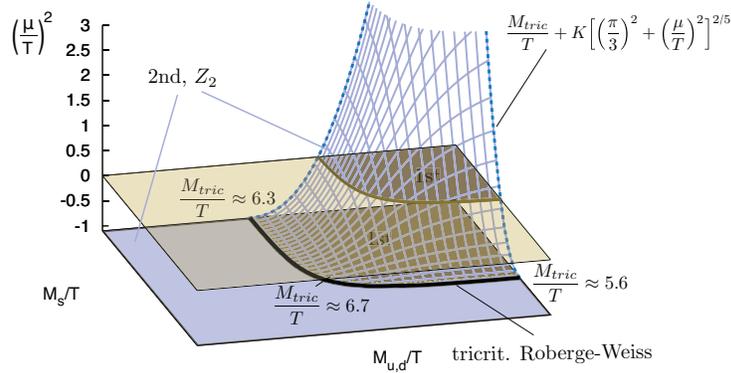}
 \caption{The extension of the top-right corner of the Columbia plot to nonzero chemical potential \cite{Fromm:2011qi}.
 }
\label{fig:Col2}
\end{center}
\end{figure}

With the inclusion of heavy quarks, the top-right corner of the Columbia plot can also be studied. The Columbia plot, see Fig.\ \ref{fig:Col} (left), indicates the expected quark mass dependence of the thermal deconfinement transition at vanishing chemical potential: a first order transition for light and heavy quarks, related to chiral and centre symmetry, and a crossover for  intermediate-mass quarks. An extension to nonzero chemical potential makes the Columbia plot three-dimensional, as shown in 
Fig.\ \ref{fig:Col2} for the heavy quark corner: the vertical axis indicates $(\mu/T)^2$, with negative values corresponding to imaginary $\mu=i\mu_\rmI$. Due to the interplay with the centre-symmetry, a rich phase structure emerges, with a periodicity in the imaginary $\mu$ direction with period $2\pi T/3$ \cite{Roberge:1986mm}. In the high-temperature deconfined phase this results in a Roberge-Weiss (RW) transition at $\mu_{\rm RW} = i\pi T/3$, where the expectation value of the Polyakov loop is rotated by an element of the centre and the trivial vacuum at vanishing $\mu$ is no longer preferred. The first-order transition at $\mu=\mu_{\rm RW}$ is only present at high temperature;  in the confined phase at low temperature, the periodicity is smoothly realized. 
The RW transition at fixed $\mu_\rmI$ therefore necessarily ends at an endpoint at a temperature $T_{\rm RW}$, see Fig.\ \ref{fig:Col} (right) \cite{Roberge:1986mm,deForcrand:2002ci,D'Elia:2002gd}.

For light and heavy quarks, the thermal deconfinement transition is first order. It is expected that this transition is connected with the RW transition at imaginary $\mu$, leading to the phase diagram in Fig.\ \ref{fig:Col} (right). The RW endpoint is then a triple point, where three first-order lines meet. For quarks with an intermediate mass, the deconfinement transition is a crossover and the RW endpoint is a second-order point at the end of a first-order line. This has been confirmed with lattice QCD simulations in the two-flavour \cite{D'Elia:2009qz} and the three-flavour \cite{deForcrand:2010he} case (the triple point has also been seen with holographic methods \cite{Aarts:2010ky}). Returning to the Columbia plot in Fig.\  \ref{fig:Col2}, these results imply that the plane at fixed $(\mu/T)^2 = -(\pi/3)^2$ is critical, with a tricritical line separating the first-order region for heavy quarks from the second-order region for intermediate-mass quarks.
Moving away from the RW plane, the tricritical line turns into a critical (second-order) surface, which separates the thermal first-order and crossover transitions.
An intriguing recent insight is that the tricritical line plays an important role in determining the curvature of this critical  surface at $(\mu/T)^2 > -(\pi/3)^2$, via tricritical scaling \cite{deForcrand:2010he}. The shrinking of the first-order region as $(\mu/T)^2$ is increased, according to tricritical scaling, has been confirmed in the effective Polyakov loop models discussed above \cite{Fromm:2011qi} as well as in the three-state Potts model \cite{deForcrand:2010he}, simulated some time ago \cite{Kim:2005ck}.
Clearly, this goes far beyond simple analytical continuation and motivates the study of the three-dimensional Columbia plot using imaginary chemical potential also for light quarks \cite{Bonati:2012pe}.

\section{Into the complex plane!}
\label{sec:complex}

\begin{figure}[h]
\vspace*{0.5cm}
 \begin{center}
  \includegraphics[height=4.2cm]{plot-rho.eps}
  $\quad$
    \includegraphics[height=4.2cm]{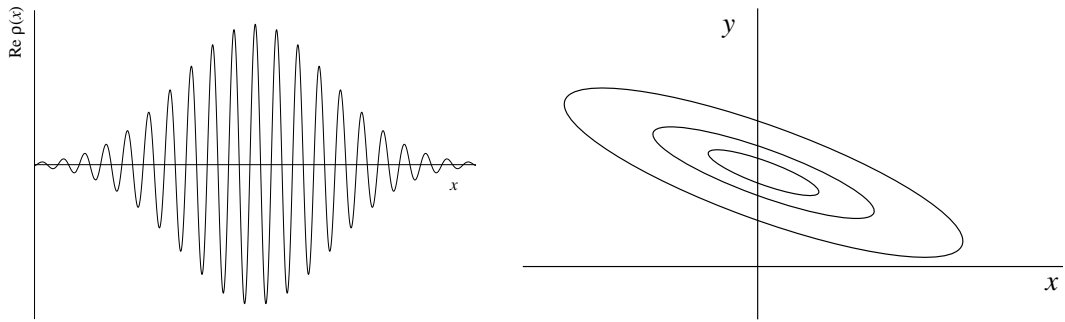}
 \caption{Weight in the partition function: while the original weight  $\rho(x)$ on the real axis is complex and oscillatory (left), there may be a real and positive weight $P(x,y)$ in the complex plane (right).}
\label{fig:rho}
\end{center}
\end{figure}

For the remainder of this overview, let us take a few steps back and try to tackle the problem of a complex Boltzmann weight head on. Consider first the simple integral
\be
Z = \int_{-\infty}^\infty dx\, e^{-S(x)}, 
\quad\quad\quad
S(x) = \half ax^2+ibx.
\ee
In Kindergarten one used to be taught to do this integral by completing the square (or by a saddle point approximation, for more complicated functions), with the result that the variable is taken into the complex plane, $x\to x-ib/a$.
The lesson to be drawn from this is that one should not insist on staying on the real axis, but be more imaginative.
This leads to a radically different approach: all degrees of freedom are complexified, $x\to z=x+iy$, yielding an enlarged configuration space with literally new directions to explore.
In SU($N$) gauge theories or matrix models, this extension implies that the dynamics takes place in SL($N, \mathbb{C}$) \cite{Berges:2006xc,Aarts:2008rr}. 
 The hope is that a real and positive distribution exists in this complexified space, as illustrated in Fig.\ \ref{fig:rho}.

The question is of course how to find this distribution. Before discussing complex Langevin dynamics, I briefly mention a recent proposal for studying high-density QCD on a Lefschetz thimble \cite{Cristoforetti:2012su}. The idea here is to deform the integration contour into the complex plane and associate (real) integration domains with each stationary point, the so-called thimbles. The original path integral is then expressed as a sum over integrals over thimbles.  This proposal, partly motivated by ideas due to Witten and Morse theory, is currently being implemented and results, first for cases simpler than QCD, will be available soon \cite{Cristoforetti:2012uv}.

One potential way to obtain a real and positive probability distribution $P(x,y)$ in the complexified space is via the solution of a stochastic process. This is exactly the aim of complex Langevin (CL) dynamics, proposed many years ago by Parisi \cite{Parisi:1984cs} and by Klauder \cite{Klauder:1983} and the topic of the second half of this contribution. Since CL has been around for 30 years, the natural question to ask is why to talk about it here. Moreover, while for real actions the procedure is equivalent to path integral quantization \cite{Damgaard:1987rr} and known as stochastic quantization \cite{Parisi:1980ys}, for complex actions there is no formal proof of correctness and ``disasters of various degrees'' were encountered by e.g.\ Ambj{\o}rn and others in the mid-1980s \cite{Ambjorn:1985iw,Ambjorn:1986fz}.
However, in recent years examples in field theory  have been given for which CL can solve severe sign and Silver Blaze problems and (importantly!)  also gives the correct result. In addition, our analytical understanding has improved steadily. In the literature various scattered results can be found, starting from the mid-1980s; here I will review mostly results obtained from 2008 onwards in the context of finite density \cite{Aarts:2008rr} .

\subsection{Sign and Silver Blaze problems}

Let us first consider two examples where CL has successfully solved a severe sign problem as well as a Silver Blaze problem: the four-dimensional weakly interacting relativistic Bose gas \cite{Aarts:2008wh,Aarts:2009hn} and one-dimensional QCD \cite{arXiv:1006.0332}, both at nonzero chemical potential.

\begin{figure}[h]
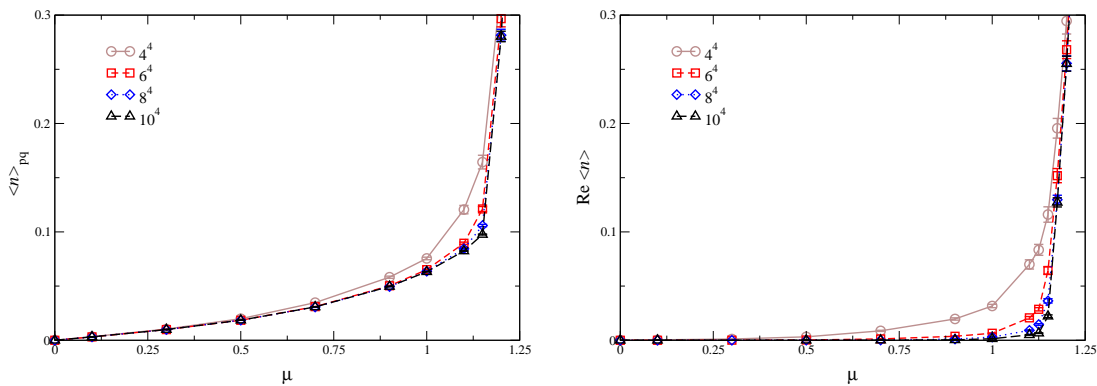

  \begin{center}
    \includegraphics[height=5cm]{plot_phq_dens_4-6-8-10_m1_l1_v2.eps}
        $\quad$
    \includegraphics[height=5cm]{plot_redens_4-6-8-10_m1_l1_v4.eps}
  \caption{Density $\bra n\ket$ in the relativistic Bose gas as a function of $\mu$ in the phase-quenched  (left) and the full theory (right), for four different lattice volumes (with $am=\lambda=1$). The onset occurs at $\mu_c\sim 1.15$ \cite{Aarts:2008wh}.}
  \label{fig:bose1}
\end{center}
\end{figure}

As in the case of QCD, the continuum and lattice actions for the Bose gas satisfy the usual symmetry, $S^*(\mu) = S(-\mu^*)$. For instance, the continuum action is
\be
S = \int d^4x \, \left[    |\partial_\nu \phi|^2+ (m^2-\mu^2)|\phi|^2 
+ 
\mu\left(\phi^*\partial_4\phi - \partial_4\phi^* \phi \right)
+ \lambda |\phi|^4\right].
\ee
 At zero temperature, the theory has a Silver Blaze problem: in the full theory onset occurs at $\mu_c\sim m$, with 
 $\mu$-independence when $\mu<\mu_c$. On the other hand, in the phase-quenched theory the term linear in $\mu$ is absent, leaving just an effective $\mu$-dependent mass term, $m^2_{\rm eff} = m^2-\mu^2$, and there is immediate $\mu$-dependence as $\mu$ is increased. The purely imaginary term, linear in $\mu$, should cancel this $\mu$-dependence in the full theory (the same considerations hold of course for the lattice discretized theory) \cite{Aarts:2008wh}. This is illustrated in Fig.\ \ref{fig:bose1}. Shown are the densities $\bra n\ket$ in the phase-quenched (left) and the full (right) theory, obtained with real resp.\ complex Langevin dynamics, on four lattice volumes from $4^4$ to $10^4$ (lattice units are used, with $m = \lambda=1$).
   In the full theory, the density is approaching zero in the thermodynamic and zero-temperature limit, until the transition to a condensed phase occurs at $\mu\sim 1.15$, in agreement with mean-field expectations \cite{Aarts:2009hn}.
 In the phase-quenched theory the density increases immediately, again in agreement with mean-field expectations. 
The effects of the complex  phase, $e^{i\theta}=e^{-S}/|e^{-S}|$, are therefore correctly incorporated by complex Langevin dynamics. Furthermore the thermodynamic limit poses no~problem.

\begin{figure}[t]
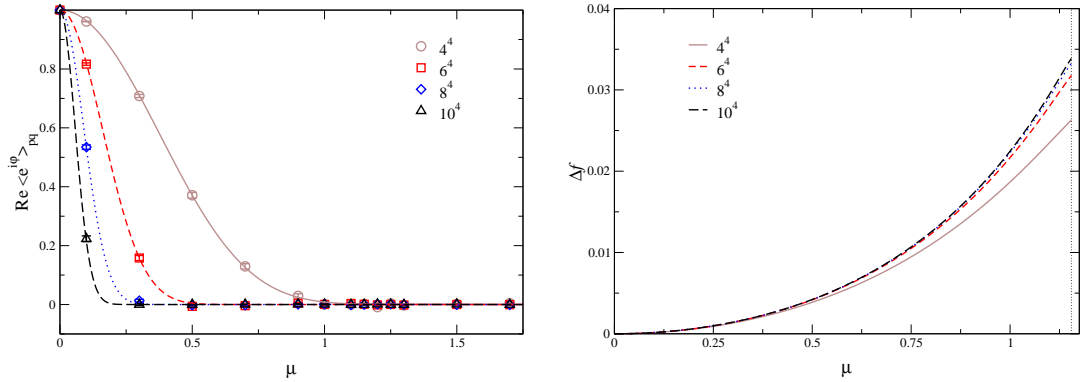

  \begin{center}
    \includegraphics[height=5cm]{plot_phq_cosf_4-6-8-10_m1_l1_MFv2.eps}
   $\quad$
   \includegraphics[height=5cm]{ana_Deltaf_MF.eps}
   \caption{Average phase factor in the phase-quenched relativistic Bose gas  for four different lattice volumes, indicating the severity of the sign problem \cite{Aarts:2008wh}. The lines are mean-field predictions (left). Difference in free energy densities between the full and the phase-quenched theory,  in the mean-field approximation \cite{Aarts:2009hn} (right).
        }
  \label{fig:bose2}
\end{center}
\end{figure}

The severity of the sign problem can be studied by analyzing the average phase factor in the phase-quenched theory. Recall that 
\be 
\label{eq:df}
\bra e^{i\theta}\ket_{\rm pq}  = \frac{Z}{Z_{\rm pq}} = e^{-\Omega\Delta f},
\quad\quad\quad
\Delta f = f-f_{\rm pq},
\quad\quad\quad
\Omega = N_s^3N_\tau.
\ee
The result is shown in Fig.\ \ref{fig:bose2} (left). It can be seen that the phase factor behaves exactly as expected: it is 1 when $\mu=0$, but as soon as $\mu>0$ it decreases, in agreement with the density plots above. As the volume is increased at fixed $\mu$, the phase factor drops exponentially. Around the critical chemical potential, $\mu_c\sim 1.15$, the phase factor vanishes even on the $4^4$ volume; yet this does not affect the detection of the phase transition. It is possible to compute the difference in free energies $\Delta f$ in mean-field theory on the lattice \cite{Aarts:2009hn}. The result is shown in  Fig.\ \ref{fig:bose2} (right) in the symmetric phase. Using this in Eq.\ (\ref{eq:df}) yields the curves in  Fig.\ \ref{fig:bose2} (left), in agreement with the numerical data. We conclude that CL can solve this theory, with a severe sign and Silver Blaze problem, in the thermodynamic limit.

\begin{figure}[t]
  \begin{center}
    \includegraphics[height=5cm]{plot_Sigma_mu1_n4-10.eps}
    $\quad\quad$
     \includegraphics[height=5cm]{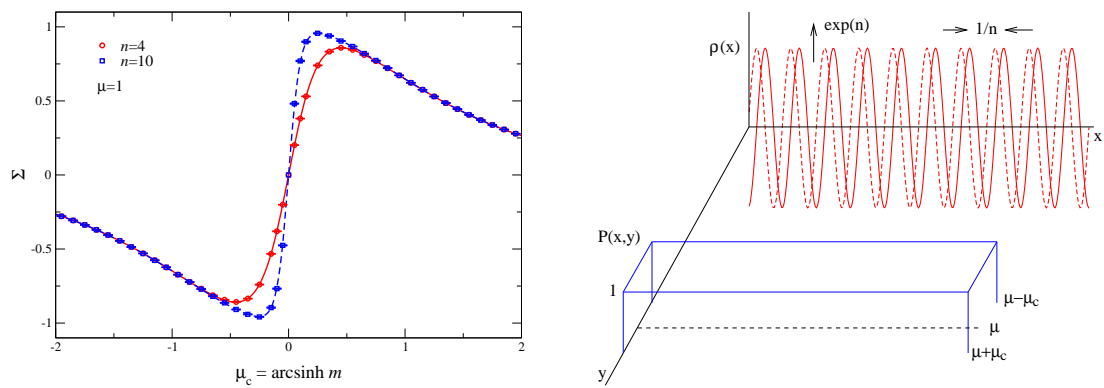}
  \caption{One-dimensional QCD: Chiral condensate as a function of the quark mass for two lattice volumes $n$ (left). Complex Langevin solution to the sign and Silver Blaze problems: the original complex and oscillatory weight $\rho(x)$ is replaced by a flat and localized distribution $P(x,y)$ in the complex plane (right) \cite{arXiv:1006.0332}. 
  }
 \label{fig:1dqcd}
 \end{center}
\end{figure}

One-dimensional QCD  \cite{Gibbs:1986xg,Bilic:1988rw,Ravagli:2007rw}
is exactly solvable and shares many features with QCD in the Silver Blaze region:  the phase-quenched theory has a transition at $\mu=\mu_c=\arcsin m$, with $m$ the quark mass, while the full theory has no transition at all, at least when the gauge group is U($N_c$). The sign problem is severe when $\mu>\mu_c$.  The chiral condensate can be written as an integral over the spectral density,
\be
\label{eq:sigma}
\Sigma = \int d^2z\, \frac{\rho(z;\mu)}{z+m},
\ee
where $\rho(z;\mu)$ is complex, oscillatory and $\mu$-dependent \cite{Ravagli:2007rw}. The condensate itself is $\mu$-independent (Silver Blaze): for this to happen, the integral (\ref{eq:sigma}) has to be carried out with great precision. When $\mu>\mu_c$, this is nontrivial: 
after writing $z=(e^{ix+\mu}-e^{-ix-\mu})/2$, the spectral density has a diverging amplitude and  vanishing period in the thermodynamic limit,  $\rho\sim e^{n(\mu-\mu_c)+inx}$, where $n$ is the one-dimensional volume. Only when all the oscillations are integrated precisely does the $\mu$-dependence disappear.
CL has no problems with this: the result for the chiral condensate is shown in Fig.\ \ref{fig:1dqcd} (left) as a function of $\mu_c$ at fixed $\mu=1$ and is seen to agree with the exact result. The discontinuity at $\mu_c=m=0$ emerges in the thermodynamic limit.

In this case, the distribution $P(x,y)$ in the complexified space can be computed analytically in the thermodynamic limit. It is surprisingly simple: a flat distribution which is nonzero only in a strip in the complexified space, $\mu-\mu_c<y<\mu+\mu_c$, see Fig.\ \ref{fig:1dqcd} (right). The complicated dynamics on the real axis has been replaced by free diffusion in the complex plane and CL finds this distribution as a solution of the stochastic process, exactly as it says on the tin \cite{arXiv:1006.0332}.

\subsection{Analytical insight}
\label{sec:52}

As mentioned above, complex Langevin dynamics has a troubled past. Problems can be divided into three groups \cite{arXiv:0912.3360}: (i) there can be numerical runaways and instabilities, (ii) the theoretical status is unclear, (iii) there may be convergence to the wrong limit. The first problem is numerical and follows from properties of the classical drift terms appearing in the CL equation.
It can be partly eliminated by choosing a small enough Langevin stepsize \cite{Berges:2005yt} and solved by introducing an adaptive stepsize \cite{arXiv:0912.0617}. This has been implemented in a variety of theories, including SU(3) gauge theory, and no instabilities have been observed. It should be mentioned that theories which suffer from extreme instabilities and runaways are also prone to converge to the wrong limit (see below), so it is likely that the two are related.

The statement that the theoretical status is unclear originates from the following:
 for real actions it is relatively straightforward to show that stochastic quantization is equivalent to path integral quantization \cite{Damgaard:1987rr}. This relies on the appearance of a semi-positive Fokker-Planck hamiltonian. Unfortunately, this path can no longer be followed when the action is complex, since the associated Fokker-Planck hamiltonian is now no longer semi-positive.
  Recently, the theoretical foundation has been clarified and necessary conditions for obtaining the correct result have been identified \cite{arXiv:0912.3360,arXiv:1101.3270}. The idea is as follows: let us consider again one degree of freedom $x$ and the weight $\rho(x)\sim e^{-S(x)}$ with $S(x)$ complex. What is wanted are expectation values of the form
\be
\bra O(x,t)\ket_\rho = \int dx\, \rho(x,t) O(x),
\ee
where $\rho(x,t)$ satisfies a (complex-valued) Fokker-Planck equation ($t$ is the Langevin time)
\be
\partial_t\rho(x,t) = \partial_x\left(\partial_x+S'(x)\right)\rho(x,t),
\ee
with stationary solution $\rho(x)\sim e^{-S(x)}$. Instead, what is solved in the CL process are expectation values of the form
\be
\bra O(x,t)\ket_P = \int dxdy\, P(x,y,t) O(x+iy),
\ee
where $P(x,y,t)$ satisfies a (real-valued) Fokker-Planck equation
\be
\label{eq:FPP}
\partial_t P(x,y,t) = \left[  \partial_x\left(\partial_x-K_x\right) - \partial_y K_y \right] P(x,y,t),
\ee
with $K_x = -\re S'$, $K_y=-\im S'$. The crucial question is whether $\bra O(x,t)\ket_P = \bra O(x,t)\ket_\rho$, possibly for all times or perhaps in the limit $t\to\infty$. The answer, which relies on the use of the Cauchy-Riemann equations and partial integration, turns out to be yes, provided some conditions are met
\cite{arXiv:0912.3360,arXiv:1101.3270}: the important one is that the distribution $P(x,y)$ drops off rapidly in the imaginary $y$ direction, such that partial integration without boundary conditions is possible. 
A complication here is that the FP equation (\ref{eq:FPP}) cannot easily be solved, even in simple models (some recent examples can be found in Refs.\ \cite{Aarts:2009hn,arXiv:0912.3360,Duncan:2012tc}). A way forward is to require that partial integration is possible for products of the form $O(x+iy)P(x,y)$, for a large enough set of operators $O(x)$. 
 This condition can then be translated into criteria for correctness, which can be written as $\bra LO(x+iy)\ket=0$, with $L$ a Langevin operator, and measured during a numerical simulation (see below).
In any case, the main lesson is that the distribution should be sufficiently localized. In the case of one-dimensional QCD discussed above, this is achieved perfectly since $P(x,y)$ vanishes outside a strip.

The third problem of convergence to the wrong limit is the hardest one. It had been observed in the early days \cite{Ambjorn:1985iw,Ambjorn:1986fz} and a more recent study in the three-dimensional XY model at nonzero chemical potential revealed an intriguing connection with the phase structure in the $\beta-\mu$ plane \cite{arXiv:1005.3468}. I will come back to this and possible resolutions towards the end.

\subsection{The SU(3) spin model revisited}

I now return to the SU(3) spin model of Eq.\ (\ref{eq:SU3}) to illustrate some of issues discussed above. While this model had been solved with CL earlier \cite{KW,BGS}, no detailed justification was provided. 
This was rectified in Ref.\ \cite{Aarts:2011zn}, in which the reliability of CL was studied using a variety of tests. Here I will present some results based on analyticity in $\mu^2$, a low-order Taylor expansion, the criteria for correctness mentioned above, and finally a comparison with the flux representation.

\begin{figure}[t]
  \begin{center}
    \includegraphics[height=5cm]{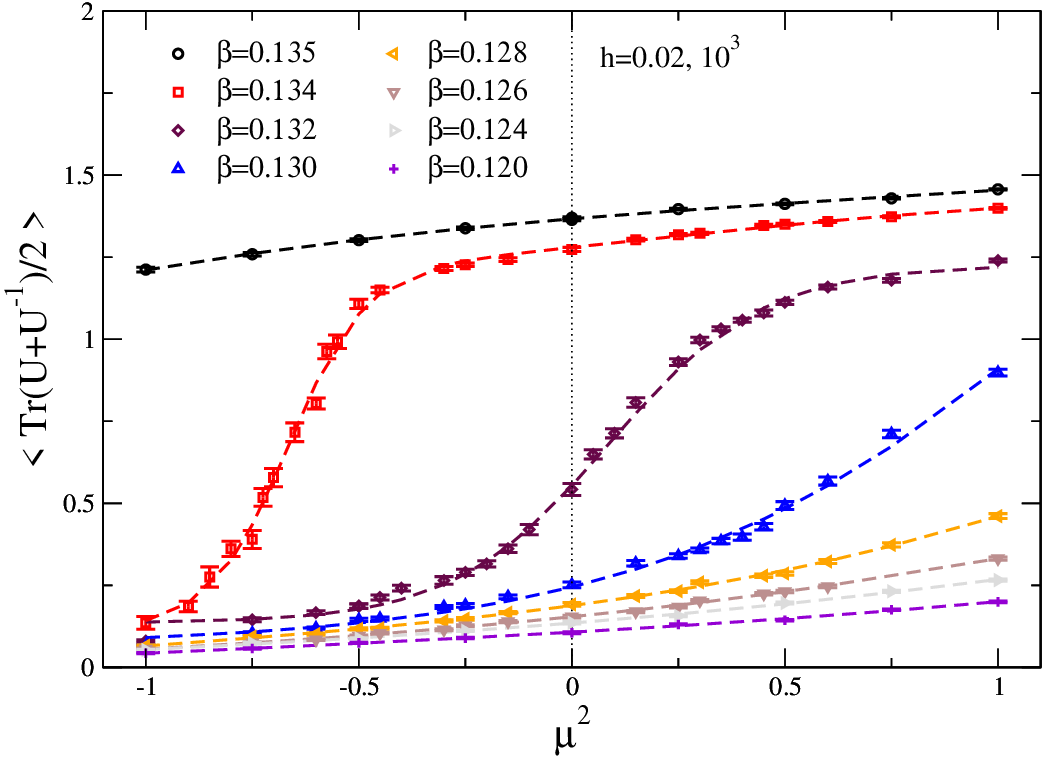}
        $\quad$
    \includegraphics[height=5cm]{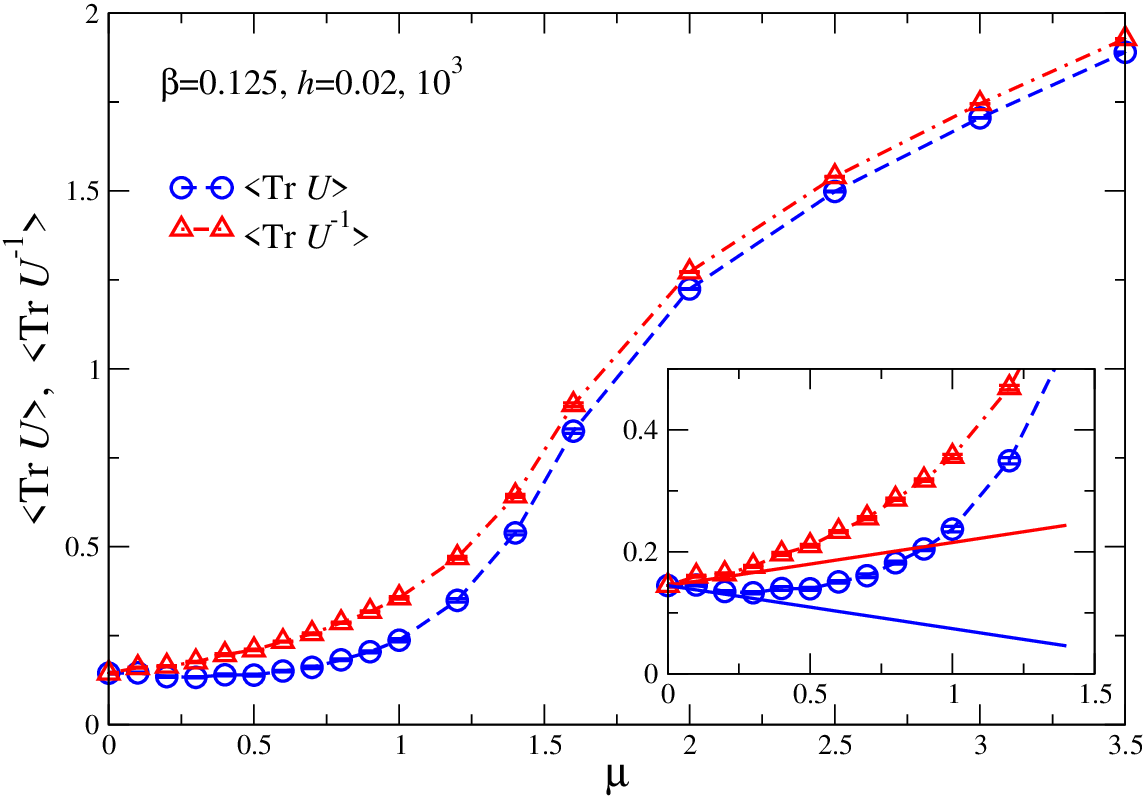}
 \caption{Analyticity in $\mu^2$ in the SU(3) spin model: $\bra \Tr (U+U^{-1})/2\ket$ as a function of $\mu^2$ for a number of $\beta$ values. Data at imaginary $\mu$ (with $\mu^2\le 0$) has been obtained with real Langevin dynamics, data at real $\mu$ (with $\mu^2> 0$)  with complex Langevin dynamics
 (left). (Conjugate) Polyakov loops $\bra \Tr U\ket$ and $\bra \Tr U^{-1}\ket$ as a function of $\mu$. The inset shows a comparison with the lowest-order Taylor expansion (right) \cite{Aarts:2011zn}.   }
\label{fig:su31}
\end{center}
\end{figure}

In Fig.\ \ref{fig:su31} (left) $\bra \Tr (U+U^{-1})/2\ket$ is shown as a function of $\mu^2$ (note that $\Tr U$ and $\Tr U^{-1}$ were earlier denoted with $P$ and $P^*$). At nonzero real  $\mu$, $\bra \Tr U\ket$ and $\bra\Tr U^{-1}\ket$ are both real but not identical, while at nonzero imaginary $\mu$ they are complex with opposite imaginary parts. The sum is always real and, since it is even under charge conjugation, a function of $\mu^2$ instead of $\mu$. 
For imaginary $\mu$ the action is real and the data has been obtained with real Langevin dynamics, for real $\mu$  complex Langevin dynamics has been used. The $\beta$ values are chosen both below and above the critical $\beta_c\sim 0.133$ at $\mu=0$. The transition gets weaker as $\mu^2$ is increased, in agreement with the Columbia plot discussion for heavy quarks. The important message from Fig.\ \ref{fig:su31} (left) is that the numerical data at $\mu^2>0$ appear to be completely consistent with those at negative $\mu^2$, where the reliability of Langevin dynamics poses no problem.
This is a first test CL should pass (and which it fails in e.g.\  the symmetric phase of the three-dimensional XY model \cite{arXiv:1005.3468}).

As a second test we study the splitting between $\bra \Tr U\ket$ and $\bra\Tr U^{-1}\ket$ at small $\mu$. This model does not have a Silver Blaze region, as can be seen from e.g.\ the expression for the density 
$\left\bra n\right\ket  = \left\bra he^\mu \Tr U - h e^{-\mu}\Tr U^{-1} \right\ket$.
It follows from the absence of the possibility to take the zero-temperature limit, since $N_\tau$ is no longer present as a free parameter.
At nonzero $\mu$, $\bra \Tr U\ket$ and $\bra\Tr U^{-1}\ket$ differ therefore immediately and behave as
\be
\bra \Tr U\ket = c_1+c_2h\mu +{\cal O}(\mu^2),
\quad\quad\quad
\bra \Tr U^{-1}\ket = c_1-c_2h\mu +{\cal O}(\mu^2),
\ee
where the coefficient of the linear term $c_2$ is determined from a Polyakov loop correlator (note that $c_2<0$) 
\be
c_2 = \frac{1}{2\Omega}\sum_{xy} \left\bra \Tr \left(U_x-U_x^{-1}\right)  \Tr \left(U_y-U_y^{-1}\right)  \right\ket_{\mu=0}.
\ee
This is illustrated in Fig.\ \ref{fig:su31} (right). The inset shows that the CL results agree with the lines predicted by the first-order Taylor expansion at very small $\mu$.

 \begin{figure}[h]
  \begin{center}
    \includegraphics[height=5cm]{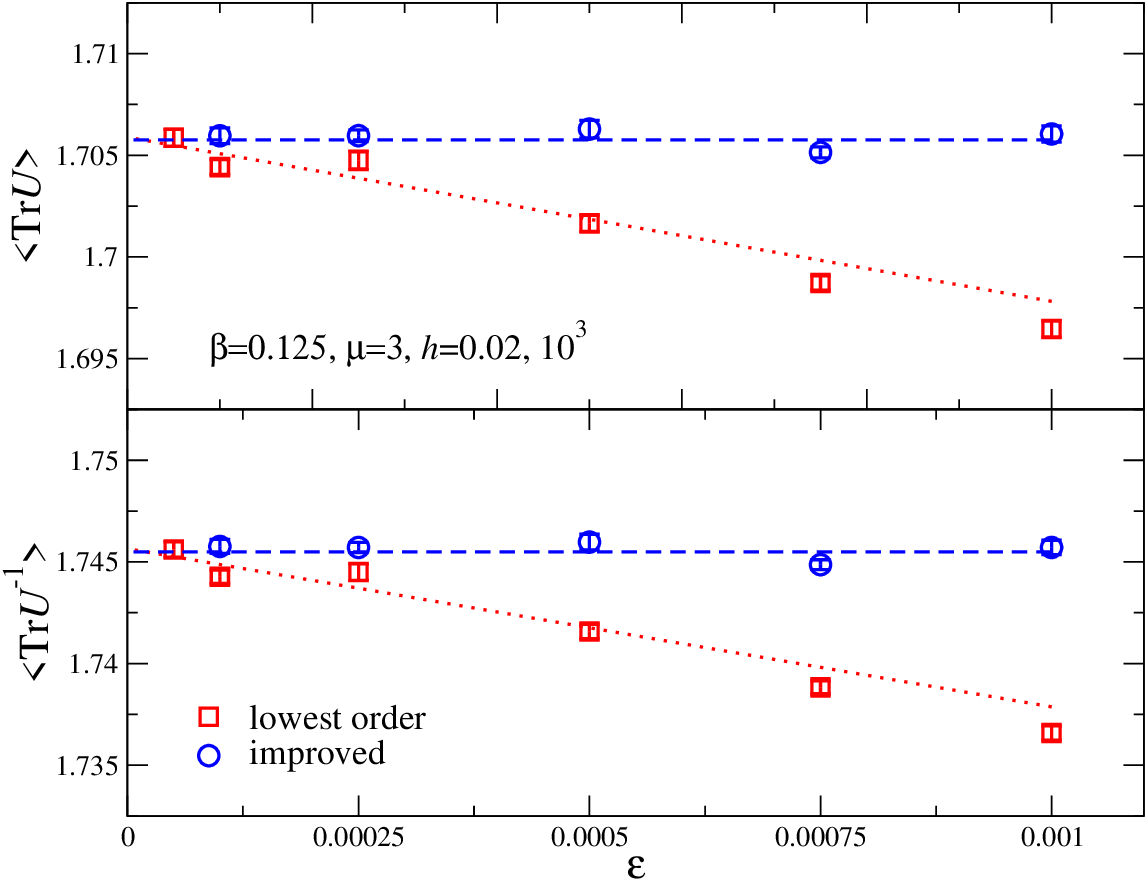}
        $\quad$
    \includegraphics[height=5cm]{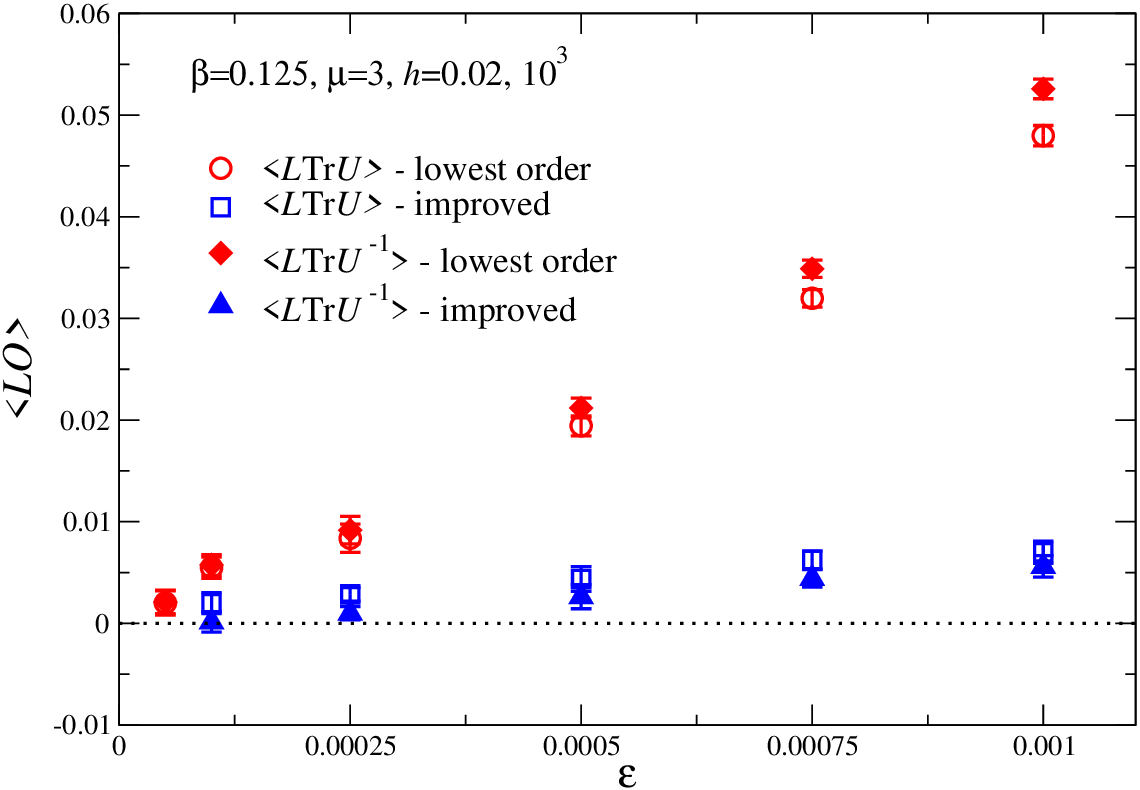}      
  \caption{
  Langevin stepsize dependence of $\bra \Tr U\ket$ and $\bra\Tr U^{-1}\ket$ (left) and 
   the real part of  $\bra L\Tr U\ket$ and $\bra L\Tr U^{-1}\ket$ (right) using the standard lowest-order (red) and the improved (blue) algorithm \cite{Aarts:2011zn}.
  }
\label{fig:su32}
\end{center}
\end{figure} 

The tests described so far use the connection with vanishing or imaginary $\mu$ and are therefore only applicable for small values of $\mu$. In order to assess larger $\mu$ values, we turn to the criteria for correctness, $\bra LO\ket=0$, with $L$ a Langevin operator and $O=\Tr U, \Tr U^{-1}$. The results are presented in Fig.\ \ref{fig:su32}: the figure on the left shows $\bra \Tr U\ket$ and $\bra \Tr U^{-1}\ket$ as a function of the Langevin stepsize $\epsilon$, while in the figure on the right the criteria are shown.
Some explanation is in order. Numerical solutions of discretized versions of the Langevin equation, with stepsize $\epsilon$, suffer from finite stepsize corrections. For the lowest-order standard discretization, these corrections are linear in $\epsilon$. This is clearly visible in Fig.\ \ref{fig:su32}, both for the observables and the criteria (red symbols). It is possible to improve this behaviour by implementing a higher-order algorithm. In Ref.\  \cite{Aarts:2011zn} we have chosen for the algorithm described in Ref.\ \cite{CCC} (for real Langevin dynamics), which is easy to implement. Analytically, one expects a cancellation of the linear term in $\epsilon$, but a remaining dependence on $\epsilon^{3/2}$, but we found effective stepsize independence in the case of the observables -- see the blue symbols in Fig.\ \ref{fig:su32} (left) -- and a much reduced stepsize dependence for the criteria -- blue symbols in Fig.\ \ref{fig:su32} (right). In the limit that $\epsilon\to 0$, both calculations agree and the criteria for correctness go to zero, as it should be.
As a general remark, we found that the criteria for correctness are very sensitive: for instance also in the case of real actions and real Langevin dynamics, they indicate the presence of finite stepsize corrections \cite{Aarts:2011sf}. We also note that the chemical potential in Fig.\ \ref{fig:su32} is not small  ($\mu=3$):  the analyticity around $\mu=0$ is then no longer of use and a justification as discussed here is necessary.

\begin{figure}[h]
  \begin{center}
    \includegraphics[height=5cm]{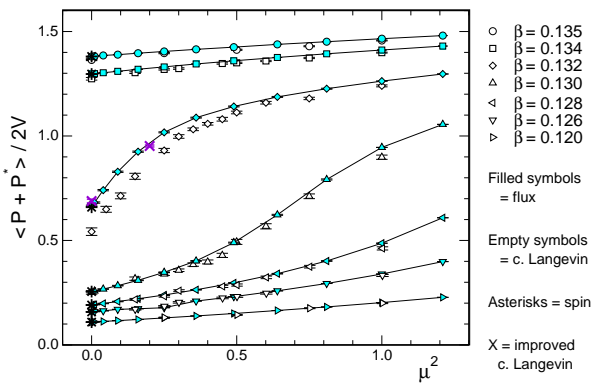}
  \caption{
   $\bra \Tr (U+U^{-1})/2\ket$ as a function of $\mu^2$, as obtained with  the flux algorithm (filled symbols), the lowest-order CL algorithm (empty symbols) and the improved CL algorithm (crosses). The values at $\mu=0$ (asterisks) have been obtained with importance sampling in the original formulation 
  \cite{Mercado:2012ue}.  
  }
\label{fig:su33}
\end{center}
\end{figure} 

A final justification comes from a comparison with the flux representation discussed above. In Fig.\ \ref{fig:su33} the $\mu^2>0$ side of Fig.\  \ref{fig:su31} (left) is shown again, but in this case there are also symbols representing the results obtained with the flux formalism, which is sign-problem free. Overall, agreement is seen, except in the critical region at $\beta=0.132$. Interestingly, the discrepancy is also present at $\mu=0$, where the action is real, which suggests that it is not an issue related specifically to {\em complex} Langevin dynamics. Instead, the problem lies with the finite Langevin stepsize, since 
the Langevin results in Figs.\ \ref{fig:su31} and  \ref{fig:su33} have been obtained using the lowest-order algorithm without an extrapolation to zero stepsize. This can be remedied with the improved algorithm, and indeed the data indicated with the blue crosses is seen to agree with flux representation.
This agreement is also present at larger $\mu=3$ and in both the ordered and disordered phases \cite{Mercado:2012ue}.  The conclusion is that CL is seen to work in the three-dimensional SU(3) spin model in the entire phase diagram.

\subsection{Modifications of the complex Langevin process}

In the SU(3) spin model, CL was seen to pass all the tests. The important question is to understand why. The crucial observation here is that the distribution in the complexified field space turns out to be sufficiently localized and that there is fast decay in the imaginary direction, as required by the mathematical justification discussed in Sec.\ \ref{sec:52}. In contrast, in the disordered phase of the XY model the distribution is broad -- there is slow decay -- and CL fails \cite{arXiv:1005.3468}. It turns out that the relevant feature that distinguishes the XY model from the SU(3) model is that the latter is nonabelian, with a nontrivial Haar measure. It is exactly the presence of the reduced Haar measure in the spin model that ensures that the complexified field space is explored in a controlled fashion, with a localized distribution, and that the real manifold is stable under small fluctuations, since its contribution to the drift is purely restoring \cite{Aarts:2012ft}.

\begin{figure}[b]
  \begin{center}
    \includegraphics[height=5cm]{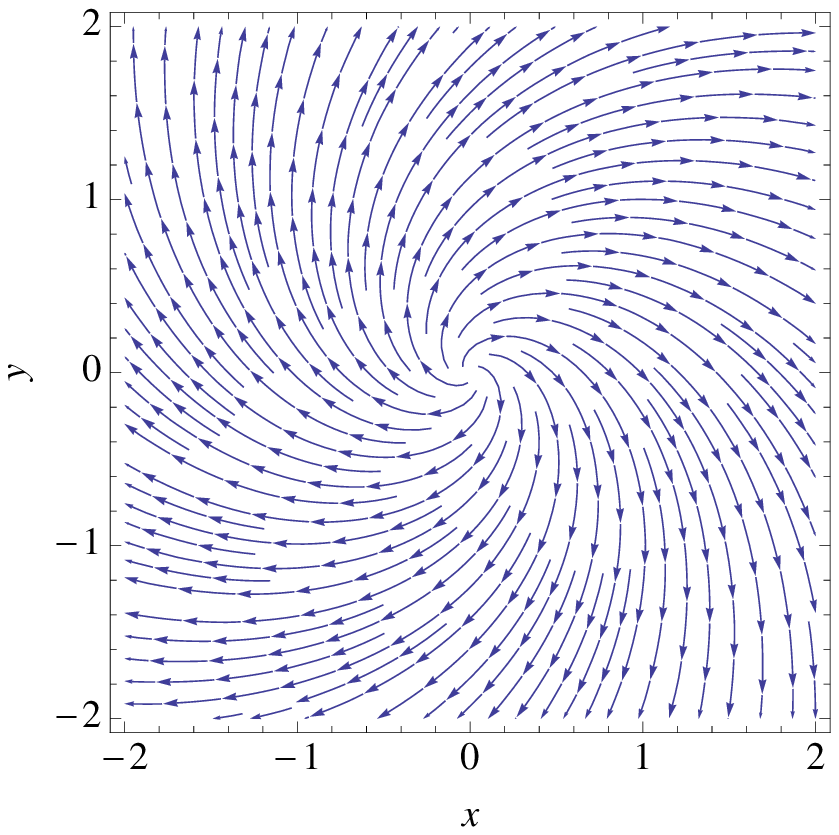}
  $\quad\quad\quad$
       \includegraphics[height=5cm]{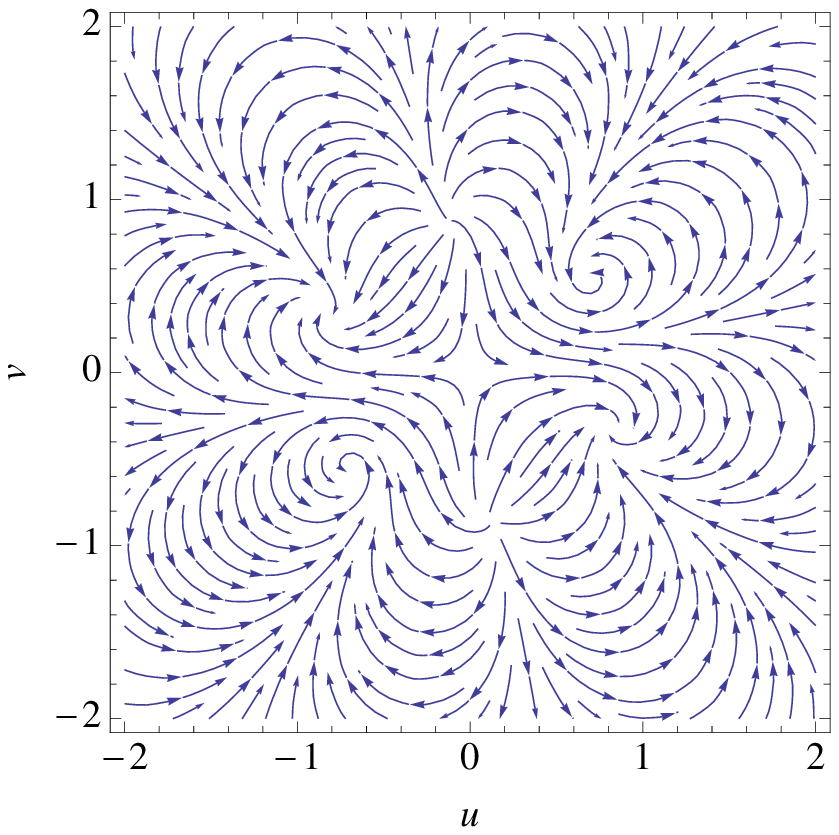}
  \caption{Classical flow diagram in the Gaussian model with 
$\sigma=-1+i$ in the original formulation (left) and after the transformation $x=u^3$ (right)
\cite{Aarts:2012ft}.
}
\label{fig:jac}
\end{center}
\end{figure}

The question is whether this insight can be employed in a practical manner to improve wrongly converging processes.
 Since the reduced Haar measure is essentially a Jacobian, one may consider generating a nontrivial Jacobian by a field redefinition, chosen in such a way that the resulting force is restoring in the complexified space. Consider an action $S(x)$ with a drift term $K(x) = -S'(x)$, which happens to lead to unstable or wrongly converging dynamics after complexification, $x\to x+iy$. A field redefinition, $x=x(u)$, will then lead to  a new process, with the action $S_{\rm eff}(u) = S(x(u)) - \ln J(u)$, where $J(u)=\partial x(u)/\partial u$ is the Jacobian. The corresponding drift, $K(u) = -S_{\rm eff}'(u) = -S'(u)+J'(u)/J(u)$, is singular at $J(u)=0$, but, importantly, may lead to a very different Langevin process after complexification, $u\to u+iv$.

 A simple and amusing example to illustrate how this works is the Gaussian integral,
 \be
 Z = \int_{-\infty}^\infty dx\, e^{-\half\sigma x^2}, 
 \quad\quad\quad
 \sigma=a+ib,
 \quad\quad\quad
 \bra x^2\ket = \frac{1}{\sigma} = \frac{a-ib}{a^2+b^2},
 \ee
but with a negative real coefficient $a<0$. After complexification, the classical drift is highly unstable as illustrated in Fig.\ \ref{fig:jac} (left) and the complex Langevin process does not converge. A simple redefinition $x(u)=u^3$ changes the flow diagram dramatically.  After complexification $u\to u+iv$, the contribution to the drift from the Jacobian, $J'(u)/J(u)=2/u$, is always directed towards the real axis, stabilizing the flow, and new attractive and repulsive fixed points appear, as shown in Fig.\ \ref{fig:jac} (right) \cite{Aarts:2012ft}. 

\begin{figure}[t]
  \begin{center}
       \includegraphics[height=5cm]{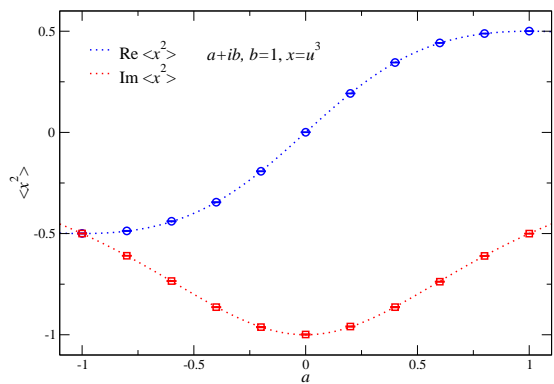}
  \caption{Complex Langevin results for the  correlator $\bra x^2\ket$ in the Gaussian model with $\sigma=a+i$ as a function of $a$ after the transformation $x=u^3$. The lines indicate the expected, analytically continued results \cite{Aarts:2012ft}.
}
\label{fig:jac2}
\end{center}
\end{figure}

Perhaps surprisingly, a CL computation of $\bra x^2\ket = \bra u^6\ket$ in the new variables yields the correct result (in the sense of being the analytically continued answer to negative $a$), see Fig.\ \ref{fig:jac2}.  A field redefinition yields therefore a process which is manifestly distinct from the original one.  As discussed in Ref.\  \cite{Aarts:2012ft}, this setup turns out to be related to the introduction of kernels in CL \cite{okamoto}.

It is possible to take these ideas a step further in SU($N$) gauge theories and modify the Langevin process by interspersing dynamical updates with SL($N, \mathbb{C}$) gauge cooling \cite{Seiler:2012wz}, with the aim of controlling the distribution in the extended field space by keeping the links as close as possible to being unitary. Exploiting the gauge freedom was earlier used in a slightly different way in an SU(2) toy model in Ref.\ \cite{Berges:2007nr}.
Very recent results in QCD with heavy quarks indicate that this is a successful strategy, allowing for the first time access to high densities in SU(3) gauge theory, although there are still some open questions about the applicability at smaller $\beta$ values \cite{Seiler:2012wz}.

\section{Summary and outlook}
\label{sec:conc}

From what is described above, I conclude optimistically that the heavy-quark corner will be settled, using a combination of techniques, formulations and effective models. As we have seen in the past years,  the interaction between conventional techniques, approaches based on new formulations which are free of the sign problem, and complex Langevin dynamics has been very stimulating and resulted in new insights and solutions to old problems. On the other hand,  in my view light quarks  at nonzero density remain a challenge. The reason is that there is a strong sensitivity to quark masses and that effective models are much less predictive.

For the case of complex Langevin dynamics, we have seen that the method can handle severe sign and Silver Blaze problems, phase transitions and the thermodynamic limit. However, a correct outcome is not guaranteed and there may be convergence to the wrong result. In the past years we have developed a better understanding, both on a formal level and with regards to numerical simulations, and developed a series of tests to justify (or dismiss) the results from numerical solutions. Importantly, this theoretical framework is also applicable to SU(3) lattice theory.

Very recently, several proposals to modify the complex Langevin process have been put forward, of which gauge cooling \cite{Seiler:2012wz} is the most promising. It will be interesting to investigate this further.

\acknowledgments

It is a pleasure to thank  Ion-Olimpiu Stamatescu, Erhard Seiler,  Frank James, Denes Sexty, Kim Splittorff and Jan Pawlowski  for collaboration and shared insight.
I also thank Philippe de Forcrand, Owe Philipsen, Christof Gattringer and Simon Hands for discussion.
Part of the work described here is carried as part of the UKQCD collaboration and the DiRAC Facility jointly funded by STFC, the Large Facilities Capital Fund of BIS and Swansea University. This work is  supported by STFC.

\end{document}